\documentclass[xcolor=dvipsnames]{cernrep} 

\usepackage{texnames}
\usepackage[T1]{fontenc}
\usepackage[bookmarks, colorlinks=true, linktoc=page, pdftex, linkcolor=black, citecolor=black, urlcolor=blue]{hyperref}
\usepackage{acronym}
\usepackage{setspace}
\usepackage{siunitx}
\usepackage{xfrac}
\usepackage{relsize}
\usepackage{subfigure}
\usepackage{colortbl}
\usepackage{sidecap}
\usepackage{mathtools}

\RequirePackage[svgnames]{xcolor}

\NewDocumentCommand\DeclareNewQuantity{mmm}{%
  	\DeclareSIUnit{#2}{#3}%
  	\DeclareDocumentCommand{#1}{O{}m}{\SI[##1]{##2}{#2}}%
	}
\DeclareNewQuantity
  \MeV
  \MeV
  {\mega\electronvolt}
  
\DeclareNewQuantity
  \pC
  \pC
  {\pico\coulomb}
  
\DeclareNewQuantity
  \keV
  \keV
  {\kilo\electronvolt}
  
\DeclareNewQuantity
  \mbar
  \mbar
  {\milli\bar}

\usepackage{fancyhdr}
\fancyhfoffset{4 mm}
\fancypagestyle{ARTTITLE}{%
\fancyhf{} 
\lhead{\footnotesize{Proceedings of the 2019 CERN--Accelerator--School course on
\it{ High Gradient Wakefield Accelerators}, Sesimbra, (Portugal)}}
\lfoot{Available online at \url{https://cas.web.cern.ch/previous-schools}}
\rfoot{\thepage\hspace*{3mm}}

   }
\begin{document}

\title{Physics of Laser-Wakefield Accelerators (LWFA)}
 
\author {Johannes Wenz, Stefan Karsch}

\institute{Centre for Advanced Laser Applications, Ludwig-Maximilians-Universit\"at M\"unchen, Munich, Germany}

\begin{abstract}
Intense ultrashort laser pulses propagating through an underdense plasma are able to drive relativistic plasma waves, creating accelerating structures with extreme gradients. These structures represent a new type of compact sources for generating ultrarelativistic, ultrashort electron beams. This chapter covers the theoretical background behind the process of LWFA. Starting from the~basic description of electromagnetic waves and their interaction with particles, the main aspects of the LWFA are presented. These include the excitation of plasma waves, description of the acceleration phase and injection mechanisms. These considerations are concluded by a discussion of the fundamental limits on the energy gain and scaling laws.
\end{abstract}


\maketitle 
\thispagestyle{ARTTITLE}



\section{Introduction}

The development of high power laser systems has enabled new approaches for generating relativistic electron beams. In a Laser-Plasma Accelerator (LPA) a transient accelerating field configuration is formed when a high intensity laser pulse propagates through a plasma. Its ponderomotive force, caused by its strong intensity gradients, deflects the electrons from their equilibrium position while the ions remain at rest. Their space-charge force causes the electrons to oscillate and constitute an electron density wave with strong longitudinal electric fields. Since the phase velocity of this plasma wave matches the group velocity of the laser pulse in the plasma, trapped electrons stay in phase with this wave for an extended period of time and can be effectively accelerated to ultra-relativistic energies. This wave structure (termed ``wakefield'') exhibits up to four orders of magnitude higher acceleration gradients than those achieved in conventional radio-frequency accelerators, thus drastically reducing the acceleration distance to reach a given energy. Not only the reduction in size, and consequently, cost of LPAs are promising, but also their unique properties regarding such important quantities as electron bunch duration, current density and emittance. These and other key features have raised global interest in exploring and developing laser-based sources as a powerful tool for medical, industrial and scientific applications.\\
The concept of a LPA in the LWFA regime was proposed over 35 years ago by \cite{bib:Tajima1979}. In their pioneering work they predicted that in a plasma, the strong transverse oscillating fields of an intense laser beam can be efficiently ``rectified'' into longitudinal fields, capable of electron acceleration to $\si{GeV}$ energies within centimeters of plasma.  
The first LPA experiments were performed with rather long laser pulses of nanoseconds to picoseconds by the plasma beat-wave accelerator scheme (PBWA) by \cite{bib:Clayton1993} and in the~self-modulated LWFA regime by \cite{bib:Modena1995}. With the emergence of ultrashort high-intensity laser systems, the~acceleration gradient could be improved from few \si{\giga\electronvolt\per\meter} up to several hundred \si{\giga\electronvolt\per\meter}, accompanied by a decrease in the transverse electron beam divergence in the so-called direct laser acceleration (DLA) or forced laser-wakefield  (FLWF) regime \cite{bib:Gahn1999, bib:Malka2002}. All of these early LPA schemes had in common an exponential electron energy spectrum with only a fraction of the accelerated charge contained in the high energy tail.%

A breakthrough in this respect was achieved in the year 2004, as three groups \cite{bib:Faure2004,bib:Geddes2004,bib:Mangles2004} simultaneously reported on the production of high-quality electron beams. They were characterized by a high charge (several $\SI{10}{\pico\coulomb}$), a quasi-monoenergetic spectrum around \SI{100}{\mega\electronvolt}  and a few-$\si{mrad}$ beam divergence. These results marked the transition to a nonlinear regime of electron cavitation, termed ``bubble'' or ``blow-out'' regime, that was predicted in simulations by \cite{bib:Pukhov2002}.
Two years later, the \SI{1}{\giga\electronvolt} barrier was broken by increasing the acceleration length to a few centimeters in a laser guiding discharge capillary \cite{bib:Leemans2006,bib:Karsch2007,bib:Walker2013,bib:Leemans2014}. Electrons approaching \SI{8}{\giga\electronvolt} have been recently measured with more powerful laser systems and enhanced guiding \cite{bib:Gonsalves2019}.%

Further improvements, such as a high shot-to-shot stability in terms of divergence, energy and beam pointing, were first demonstrated in a steady-state flow gas cell \cite{bib:Osterhoff2008}. Control over the injection process has led to a reduction in energy spread and additionally provided a knob for energy tuning. Among others,  successful injection schemes are colliding pulse injection \cite{bib:Faure2006,bib:Rechatin2009}, density down ramp \cite{bib:Geddes2008} and shockfront injection \cite{bib:Schmid2010,bib:Buck2013}, as well as related techniques \cite{bib:Faure2010,bib:Gonsalves2011,bib:Pollock2011}. %

Simultaneously, the longitudinal and transverse electron phase space has been investigated. Measurements have confirmed the few-femtosecond duration of typical LWFA bunches, up to date unrivaled by conventional accelerators \cite{bib:Buck2011,bib:Lundh2011,bib:Heigoldt2015}. The transverse emittance has been determined to  below 1 $\pi$ mm mrad \cite{bib:Weingartner2012} and few-fs-exposure shadowgrams of the wake have provided more insight into the acceleration process \cite{bib:Buck2011,bib:Savert2015}. A good introduction to experiments and theory of LWFA can be found in \cite{bib:Esarey2009} and into X-ray applications of LWFA bunches in \cite{bib:Corde2013}.%

Current research focuses on overcoming the dephasing limit on the maximum electron energy gain. Staging of LWFAs has been demonstrated, where an electron bunch from one LWFA has been successfully coupled into and gained energy in a second LWFA \cite{bib:Steinke2016}. Likewise, driver/witness type experiments as performed earlier with beams from conventional accelerators \cite{bib:Blumenfeld2007} are currently investigated. Here, an LWFA bunch is used as the driver of the plasma wave, thus overcoming the dephasing limits. Such experiments have shown that LWFA bunches are able to drive a plasma wave and can be effectively decelerated in millimeter size plasma targets \cite{bib:Chou2016, bib:Gilljohann2019}.

 
\section{Light-Matter Interaction}
\subsection{Fundamental Description of Light}
Electromagnetic phenomena, such as generation and propagation of electric ($\vec{E}\left(\vec{x},t\right)$) and magnetic ($\vec{B}\left(\vec{x},t\right)$) fields and their interaction with each other as well as with charges and currents, are described by Maxwell's equations. Their differential form is given by \cite{bib:Jackson2007}

\begin{subequations}
\begin{minipage}[t]{0.35\textwidth}
\setlength\jot{0.37cm}
\begin{align}
\nabla \cdot \vec{E}&=\frac{\rho}{\epsilon_0}, \label{eq:Max1}\\
        \nabla \cdot \vec{B}&=0, \label{eq:Max2}          
\end{align}
\end{minipage}%
\begin{minipage}[t]{0.55\textwidth}
\begin{align}
        \nabla \times \vec{E}&=-\frac{\partial \vec{B}}{\partial t}, \label{eq:Max3}\\
        \nabla \times \vec{B}&=\mu_0 \epsilon_0 \frac{\partial \vec{E}}{\partial t}+ \mu_0\vec{j}, \label{eq:Max4} 
\end{align}
\end{minipage}
\end{subequations} 
where $\rho$ and $\vec{j}$ are the charge and current density of the medium and $\epsilon_0$ and $\mu_0$ the vacuum permittivity and permeability, respectively. 
Sometimes it is more convenient to express these equations by a scalar potential $\phi$ and a vector potential $\vec{A}$, which can be obtained from Maxwell's equations\footnote {with the use of the vector identities $\vec{\nabla} \cdot (\nabla \times \vec{a})=0$ and $\nabla \times (\nabla\phi) =0$ in Eqs. (\ref{eq:Max2}) and (\ref{eq:Max3}) }: 
\begin{align}
	\vec{E}&=-\nabla \phi-\frac{\partial}{\partial t}\vec{A}, & \vec{B}&=\nabla \times \vec{A}. \label{eq:vectorpot}
\end{align}
These electromagnetic potentials are not uniquely defined, and different solutions can lead to the same electric and magnetic fields by a gauge transformation, which in principle amounts to specifying a value for the term $\nabla \cdot \vec{A}$. In the Coulomb gauge ($\nabla \cdot \vec{A}=0$), the two inhomogeneous Eqs. (\ref{eq:Max1}) and (\ref{eq:Max4}) yield the potential form of Maxwell's equations
\begin{align*}
\nabla^2\phi&=-\frac{\rho}{\epsilon_0}, &
\nabla^2\vec{A}-\mu_0\epsilon_0\frac{\partial^2 \vec{A}}{\partial t^2}&=-\mu_0\vec{j}+\mu_0\epsilon_0\nabla\frac{\partial \phi}{\partial t}.
\end{align*}
The scalar potential represents a solution for Poisson's equation and, once it is found, it can be used to solve the equation for the vector potential.  
In vacuum, in the absence of charges and currents, one possible solution is represented by a plane wave:
\begin{align}
\phi(\vec{x},t)&=0,	& \;\;\;\;\;\;\;\;\vec{A}(\vec{x},t)&=-\vec{A}_L \sin(\omega_L t -\vec{k}\vec{x}+\varphi_L) .     \label{eq:vecsol} 
\end{align}
The electric and magnetic fields $\vec{E}\left(\vec{x},t\right)$ and $\vec{B}\left(\vec{x},t\right)$ are derived from Eq. (\ref{eq:vectorpot}) to
\begin{align} 
	\vec{E}\left(\vec{x},t\right) &=\vec{E_L}\left(\vec{x},t\right)\cos(\omega_L t-\vec{k} \vec{x}+\varphi_L),\label{eq:planewave}  &
	\vec{B}\left(\vec{x},t\right)&=\vec{B_L}\left(\vec{x},t\right)\cos(\omega_L t-\vec{k} \vec{x}+\varphi_L).
\end{align}
Here, $\varphi_L$ is an arbitrary phase offset and $\omega_L=2\pi c/\lambda_L$ the angular frequency given by the wavelength $\lambda_L$. $\vec{E_L}$ and $\vec{B_L}$ are spatially ($\vec{x}$) and temporally ($t$) confined envelope functions, which oscillate in phase and are perpendicular to each other, $\vec{E} \bot \vec{B}$. The wave vector $\vec{k}$ and the angular frequency $\omega_L$ are connected via the dispersion relation in vacuum
\begin{equation}
\vec{k}^2=\omega_L^2/c^2 .
\label{eq.:dispersionrel}
\end{equation} 
Moreover, $\vec{E} \bot \vec{B}$ implies that $\vec{E} \bot \vec{k}$, $\vec{B} \bot \vec{k}$, $\vec{E} \bot \vec{A}$, and the amplitudes are related by $\vert \vec{A_L}\vert=c/\omega_L \vert\vec{B_L}\vert=1/\omega_L \vert\vec{E_L}\vert$, i.e., the $\vec{B}$-field component is $c$ times smaller than the $\vec{E}$-field.

\subsection{Gaussian Beam Optics}
\begin{figure}
    \subfigure[\textbf{Temporal profile} of the oscillating $\vec{E}$-field (cyan line) and its Gaussian envelope (dark blue dashed line) as well as the intensity profile (red line) for a pulse duration $t_{FWHM}=\SI{28}{fs}$.]
    {\includegraphics[width=0.47\textwidth]{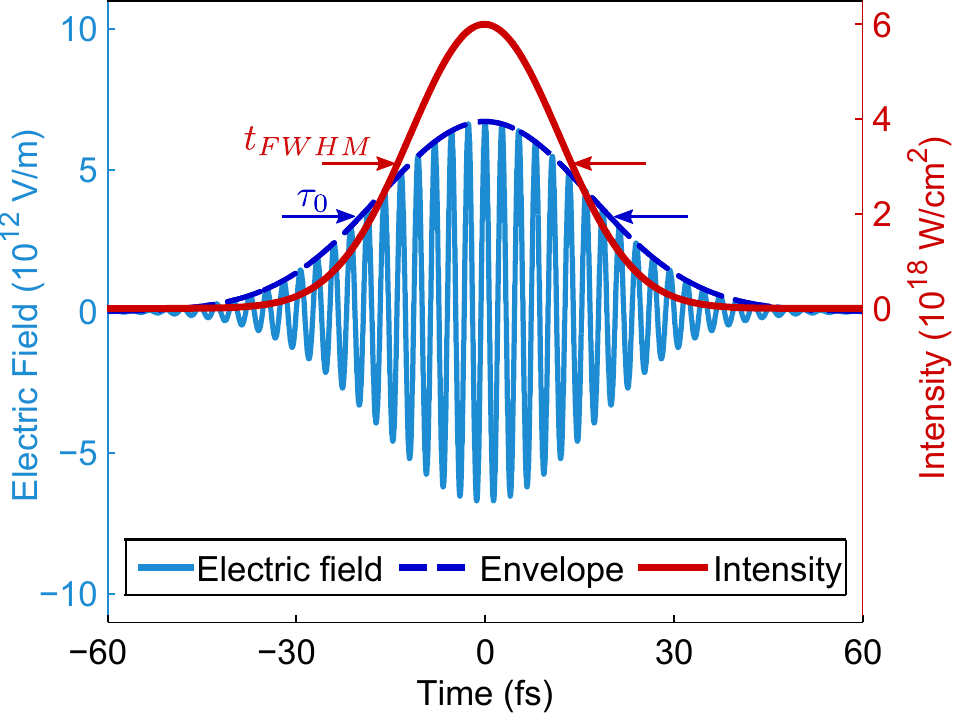}
        \label{fig:Laser_temporal_confinement}
        }
    \hspace*{0.02\textwidth}
    \subfigure[\textbf{Spatial confinement} of the ${1}/{e^2}$-irradiance (blue line) near the focal spot with $d_{FWHM}\approx\SI{22}{\micro\meter }$. The interaction region is marked in light red, defined by the intensity $I>I_0/2$.]
    {\includegraphics[width=0.47\textwidth]{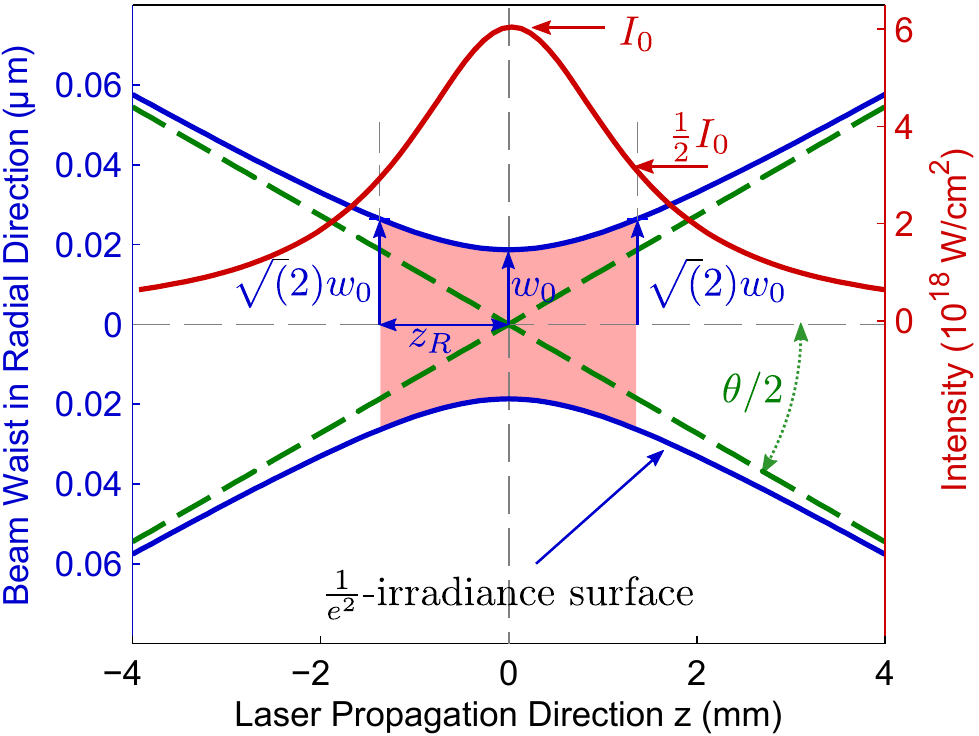}
        \label{fig:Laser_spatial_confinement}
    }
    \caption[\textbf{Spatial and temporal profile of a Gaussian laser pulse} \newline
    evaluation :	\textbackslash Th LaserPlasma\textbackslash Laser Envelope.m   \newline
    figure file a:  \textbackslash Th LaserPlasma\textbackslash Laser envelope final.pdf/svg \newline
    figure file b:	\textbackslash Th LaserPlasma\textbackslash Laser spatial confinement.pdf/svg  
]    
    {\textbf{Spatial and temporal profile of a Gaussian laser pulse} at a wavelength of $\lambda_L=\SI{800}{ nm}$, $W=\SI{1}{J}$ and $t_{FWHM}=28 fs$ focused in a $F/22$ focusing geometry.}
    \label{fig:gaussian_laser_pulse}
\end{figure}

The spatial profile of a laser pulse can be described by a Gaussian distribution. For a radial symmetric profile with radius $r$ the electric field of a monochromatic beam near the focal point can be represented as \cite{bib:Milonni2010}
\begin{equation}
\vec{E}\left(r,z,t\right)=\frac{E_0}{2}\underbrace{e^{-\frac{r^2}{w\left(z\right)^2}}}_{{\larger\textcircled{\smaller[2]1}}}
\underbrace{e^{-\frac{\left(t-z/c\right)^2}{\tau_0^2}}
}_{{\larger\textcircled{\smaller[2]2}}}
\underbrace{\cos\left(\omega_L t -k_L z +\varphi_L\right)}_{\larger\textcircled{\smaller[2]3}}\vec{e}_{pol}+ cc.,
\label{eq:gaussian_beam}
\end{equation}where the first term {\larger\textcircled{\smaller[2]1}} describes the radial envelope with the transverse 1/e radius of the beam $w(z)$. 
The~second p  art {\larger\textcircled{\smaller[2]2}} represents the temporal envelope with $\tau_0$ as the pulse duration defined at $\sfrac{1}{e}$-height of the electric field, cf. Fig. \ref{fig:Laser_temporal_confinement}. {\larger\textcircled{\smaller[2]3}} is an oscillatory term containing the carrier angular frequency $\omega_L$ and
$\vec{e}_{pol}$ is the vector describing the polarization of the laser pulse in the ($x$,$y$)-plane, i.e., $\vec{e}_{pol}=\vec{e}_{x,y}$ for linear and $\vec{e}_{pol}=(\vec{e}_x\pm i\vec{e}_y)/\sqrt{2}$ for circular polarization.%

The transverse radius of the beam evolves along the propagation direction with 
\begin{align}
w\left(z\right)=w_0\sqrt{1+\left(z/z_R\right)^2},  \;\;\;\; \;\;\;\; \text{with } 
z_R=&\pi w_0^2/\lambda_L,
\label{eq:waist_evolution}
\end{align}
where $w_0$ is the radius at $1/e$-height of the maximum electric field and $z_R$ is the \textit{Rayleigh length}, defined as the distance from the waist $w_0$ to the z-position at half the intensity. The interaction range defined as twice the Rayleigh length is called the \textit{confocal parameter}. It is marked as the shaded area in Fig.~\ref{fig:Laser_spatial_confinement}.
Far away from the focal plane, the corresponding divergence of the beam can be approximated by $\tan \theta\approx\theta\approx\Delta w(z)/ \Delta z=2\lambda_L/(\pi w_0)$. Here the origin can be considered as a point source, such that $\theta \approx D/f\equiv(F/\#)^{-1}$, where $D$ is the diameter of the collimated beam, $f$ is the focal length and $F/\#$ is the~f-number of the focusing optic.
Equating these two expressions result in the spot size. In terms of the~laser parameters, it is given by\footnote{Although the actual beam profile is rather closer to a super-gaussian beam, the use of Gaussian optics simplifies the~expressions. For a top-hat beam, the spot size is $d_{FWHM}\approx 1.03\lambda_L F/\#$.}
\begin{align*}
2w_0=&\frac{4\lambda_L f}{\pi D}.
\end{align*}
Note that the waist of the beam $w_0$ and the~pulse duration $\tau_0$ are related to the measured quantities at FWHM of the temporal and spatial intensity profile via the spot size $d_{FWHM}=\sqrt{2\ln(2)}w_0$ and the~pulse duration $t_{FWHM}=\sqrt{2\ln(2)}\tau_0$. From these quantities and the total laser pulse energy $W_L$ the laser peak power $P_0$ and the laser peak intensity $I_0=2P_0/\pi w_0^2$ can be determined. For a Gaussian pulse, as given in Eq. (\ref{eq:gaussian_beam}), they are
\begin{align}
P_0=&0.94\frac{W_L}{t_{FWHM}}, & I_0=&0.83\frac{W_L}{t_{FWHM}d^2_{FWHM}}.
\label{eq:peak_intensity}
\end{align}
On the other hand, the intensity is defined as the cycle-averaged \textit{Poynting vector} $\langle \vec{S} \rangle=\epsilon_0 c^2 \langle \vec{E} \times \vec{B} \rangle$. For a linearly polarized pulse, it depends on the electric field via
\begin{equation}
I_L= \langle \vert \vec{S} \vert\rangle_T=\epsilon_0 c   \langle \vert \vec{E} \vert ^2\rangle_T= \frac{\epsilon_0 c}{2}E_L^2. \label{eq.:intensity}
\end{equation}
The peak magnetic and electric field can be expressed in terms of $I_0$ (Eq. (\ref{eq:peak_intensity}))
\begin{align*} 
E_0=2.7\sqrt{\frac{I_0[\si{\W\per\square\cm}]}{10^{18}}}\lambda_L[\si{\micro\meter}]\times \SI{e12}{\V\per\m},  \;\;\;\; \;\;\; \;\;  B_0=0.9\sqrt{\frac{I_0[\si{\W\per\square\cm}]}{10^{18}}}\lambda_L[\si{\micro\meter}] \times \SI{e4}{\tesla\per\m}.
\end{align*}
State-of-the art high-intensity laser systems readily reach intensities beyond $\SI{e19}{\W\per\square\cm}$, corresponding to field amplitudes of $\SI{e12}{\volt\per\m}$ and $\SI{e4}{T}$. These electric fields are several order of magnitude higher than the binding fields within atoms, resulting in ionization and plasma formation in  the focal volume.

\subsection{Plane Wave Interaction with Particles} 
With the high peak intensities delivered by current state-of-the-art laser systems, even the leading edge of the laser pulse is strong enough to ionize matter nanoseconds to picoseconds before the arrival of the~peak. The intense part of the laser pulse therefore interacts with free charges.

\subsubsection{Interaction with Single Electrons}
The relativistic motion of an electron (with charge $q=-e$, mass $m_e$, velocity $\vec{v}$ and momentum $\vec{p}$) in the presence of an electromagnetic field is described by the Lorentz force \cite{bib:Gibbon2005}
\begin{equation}
\frac{d\vec{p}}{dt}=-e(\vec{E}+\vec{v_e} \times \vec{B}), \label{eq:Lorentz}
\end{equation} 
and the energy equation
\begin{equation}
\frac{d}{dt}(\gamma mc^2)=-e(\vec{v}\cdot\vec{E}).
\label{eq:energy_equation}
\end{equation}
Here, $\gamma$ is the relativistic factor and $\vec{p}$ the momentum. They are defined by 
\begin{align}
\gamma &=\sqrt{1+\frac{p^2}{m_e^2c^2}}= \frac{1}{\sqrt{1-\beta^2}}, &\vec{p}&= m_e \gamma \vec{v}, \label{eq:gammafactor}
\end{align}
where $\beta$ is the normalized velocity $\beta=\vert \vec{v_e}\vert/c$.
For a subrelativistic case ($\vec{v_e}\ll c$ and  $\vert\vec{B}\vert =  \vert\vec{E}\vert/c$) the~cross product can be neglected and the Lorentz equation simplifies for plane waves Eq. (\ref{eq:planewave}) to
\begin{equation}
m_e \frac{d\vec{v_e}}{dt}=-e\vec{E_0}\cos(\omega_Lt -\vec{k}\vec{x}+\varphi_L). \label{eq:nonrelel}
\end{equation}
Integration with respect to time yields the quiver velocity $v_e$. In the electric field $\vec{E_0}$ the electron can reach a maximum value of $v_{e,max}=e\vert\vec{E_0}\vert/m_e \omega_L$. %

A ponderomotive energy $W_{pond}$ and potential $\Phi_{pond}$ can be associated with the quiver velocity by averaging the kinetic energy over one optical light cycle.
\begin{align}
W_{pond}=&\frac{1}{2} m_e \langle \vec{v_e}^2 \rangle_T=\frac{e^2  \vec{E_0}^2}{4 m_e \omega_L^2} =e\Phi_{pond}. 
\label{eq:ponderomotive_energy}
\end{align}%

Once the quiver velocity approaches the speed of light $c$, the assumption $\vert \vec{v}\vert \ll c$ breaks down and the~$\vec{v} \times \vec{B}$ term cannot be neglected anymore. The ratio of the mean quiver energy to the electron rest mass is defined as a so-called \textit{ normalized vector potential} $a_0$. It distinguishes the subrelativistic ($a_0 \ll 1$) and the relativistic regime ($a\gtrsim 1$) 
\begin{equation}
a_0=\frac{e\vert \vec{E_0}\vert}{m_e \omega_L c}=\frac{e\vert \vec{A}\vert}{m_e c}=0.85\sqrt{I_0[\SI{e18}{W\per\square\cm}]}\cdot \lambda_L[\si{\micro\m}]. 
\label{eq:normalized_vector_potential} 
\end{equation}
For a TiSa-based laser system operating at a wavelength of $\lambda_L = \SI{800}{nm} $, $a_0=1$ is reached at laser intensities of $I_0\approx\SI{2.2e18}{W\per\square\cm}$.%

For high laser intensities, both the electric and magnetic fields are responsible for the electron motion. The electron trajectory in a plane wave consists of a drift in the z-direction with the velocity $v_D/c=a_0^2/(4+a_0^2)$ 
and Fig. 8 motion in the average rest frame of the electron, \cite{bib:Gibbon2005}. The net effect after the~transit of the laser pulse is a translation of the electron in the forward direction, but the electron is at rest again. An acceleration in an infinite plane wave is therefore not possible, a phenomenon known as Lawson-Woodward-theorem \cite{bib:Woodward1946,bib:Lawson1979}.%

However,  in a focused beam Eq. (\ref{eq:gaussian_beam}) the electron can gain energy in the following way: The strong radial intensity gradient pushes an electron, initially located on the laser axis, to an off-axis location. At this position it experiences weaker field amplitudes, and when the field reverses, the restoring force is reduced. On average the electron is pushed out from high-intensity regions by the ponderomotive force
$\vec{F_{pond}}=-e\nabla \Phi_{pond} \propto -\nabla(I_L \lambda_L^2)$.
The coupling of the laser to the electrons is mediated mainly by this quasi-force,
which can be derived from the second-order motion of electrons in the high gradients of the~light field \cite{bib:Gibbon2005}. In the linear regime it is convenient to describe $F_{pond}$ via the normalized vector potential $a$ averaged over one laser cycle, i.e., for a Gaussian pulse $a(t)= \langle a \rangle_T=a_0\exp[-(t/\tau_0)^2]$ it is given by\footnote{Note that for a linearly polarized pulse $a_0^2$ has to be replaced by $a_0^2/2$.} \cite{bib:Esarey1996}
\begin{align}
\vec{F_p}=&-m_ec^2\nabla \frac{a^2}{2}, &
\vec{F_{p,rel}}=&-m_ec^2\nabla\langle\gamma-1\rangle\propto a.
\label{eq:pond_force}
\end{align}
In the relativistic regime the ponderomotive force $F_{p,rel}$ is usually defined by the gradient of the cycle averaged electron gamma factor $\langle \gamma \rangle$. 
The associated maximal relativistic ponderomotive potential for an~electron initially at rest can be estimated in a linearly polarized wave by \footnote{from Eq. (\ref{eq:gammafactor}) for $p/m_ec=u_\perp$ as given in Eq. (\ref{eq:vectorpotential_perp})}
\begin{equation*}
\Phi_{p,rel}=\frac{m_ec^2}{e}\frac{a_0}{\sqrt{2}}.
\end{equation*} 
For example, at intensities of $\SI{8.8e18}{\W\per\square\cm}$ ($a_0=2$), the ponderomotive energy is $e\Phi_{p,rel}\approx \MeV{0.7}$. Therefore, a direct laser acceleration of electrons with $W_{el}>\MeV{10}$ using table top systems is suppressed. However, the corresponding ponderomotive force can excite a~plasma wave, which acts as intermediary and transforms the strong transversal fields of the laser into longitudinal fields suitable for acceleration. The process of longitudinal acceleration is examined below.\\

\section{Laser-Driven Plasma Waves}
As discussed in the previous section the leading edge of the laser pulse ionizes the target medium, and the intense part of the laser interacts with a plasma. Its ponderomotive force can then displace electrons from their equilibrium positions and cause macroscopic regions dominated by space charge. In order to understand these dynamics, it is necessary to consider the propagation of the electron-magnetic wave of the laser in a plasma medium.

\subsection{Electromagnetic Waves in Plasma}
As it is impossible to treat each particle individually, the motion of electrons driven by an electromagnetic wave in a plasma can be derived from a set of Eqs. (\ref{eq:Lorentzplasma}) - (\ref{eq:poisson}). It consists of the Lorentz equation, 
the~continuity equation\footnote{derived from Maxwell's Eq. (\ref{eq:Max1}) and (\ref{eq:Max4}) using the vector identity $\nabla (\nabla \times \vec{a}=0)$)} 
and Poisson's equation\footnote{a direct consequence of Maxwell's equation (\ref{eq:Max1}) and Eq. (\ref{eq:vectorpot}) in the Coulomb gauge ($\nabla A=0$).} 

\begin{align}
&\textbf{Lorentz equation:} &\frac{d \vec{p}}{d t}=\left(\frac{\partial }{\partial t} +\vec{v}\cdot \nabla \right)\vec{p}=-e[\vec{E}+\vec{v}\times\vec{B}], \label{eq:Lorentzplasma}
\\
&\textbf{Continuity equation:} &\frac{\partial n_e}{\partial t}+\nabla (n_e \vec{v})=0, 
\label{eq:cont.eq}
\\
&\textbf{Poisson's equation:} &\nabla^2 \Phi = -\frac{\rho}{\epsilon_0}=e\frac{\delta n_e}{\epsilon_0}, \label{eq:poisson}
\end{align}
where $\vec{p}=m_e\gamma\vec{v}$ Eq. (\ref{eq:gammafactor}) is the momentum and $\delta n_e=n_e-n_{e,0}$ is the local density perturbation, $\vec{j}=-e n_e \vec{v}$ the current density and $\rho=-e \delta n_e$ is the charge density, respectively.

\subsubsection{Dispersion Relation in Plasma}
If the electrons forming the plasma are expelled from their equilibrium position via the ponderomotive force, they will be pulled back by the positive ions and oscillate around their initial position with a~characteristic frequency. This oscillation frequency can be derived from Eq. (\ref{eq:Lorentzplasma}) considering small amplitudes in a cold plasma, where the initial thermal energy of the electrons is ignored. The equation of motion for the plasma fluid is given by $n_{e,0} m_e \partial \vec{v}/\partial t=-n_{e,0} e \vec{E}$, where the quadratic terms have been neglected.

It can be reformulated with Eq. (\ref{eq:Max4}) and the vector identity  $\nabla\times(\nabla\times \vec{a})=\nabla(\nabla \cdot \vec{a})-\nabla^2 \vec{a}$ to
\begin{equation*}
-\nabla(\nabla \vec{E})+\nabla^2 \vec{E}=\mu_0 n_{e,0} \frac{e^2}{m_e}\vec{E}+\mu_0 \epsilon_0 \frac{\partial^2 \vec{E}}{\partial^2t}.
\end{equation*}
For oscillating electromagnetic plane waves of the type $E\propto e^{i{k_Lx-\omega_L t}}$ propagating in a uniform medium ($\nabla \vec{E}=0$) the above equation results in the \textit{dispersion relation of a cold plasma}  
\begin{align}
\omega_L^2=\omega_p^2+c^2k_L^2 \;\;\;\;\;\;\;\;\text{with }  \omega_p\equiv\sqrt{\frac{n_{e,0} e^2}{m_e \epsilon_0}},\label{eq:dispersion_plasma}
\end{align}
where $\omega_p$ is the so called plasma frequency. For $\omega_L=\omega_p$ a critical density can be defined 
\begin{align*}
n_{e,c}\equiv&\frac{\epsilon_0 m_e}{e^2} \omega_L^2, & n_{e,c}=&\frac{1.1  }{\lambda_L^2[\si{\micro\meter}]}\SI{e21}{\per\cubic\centi\meter}.
\end{align*}
In plasmas with densities $n_{e}\!\!>\!\!n_{e,c}$ ($\omega_p\!\!>\!\!\omega_L$), the characteristic time scale of the plasma is shorter than the optical period of the light wave and the electrons can follow the field oscillation, shield the inside of the medium and the laser is reflected. The medium is referred to as overdense. In contrast, the plasma is called underdense for plasma densities $n_{e}\!\!<\!\!n_{e,c}$. and the electrons cannot follow the incoming radiation, allowing a propagation of the wave. The corresponding phase and group velocities $v_{ph}$ and $v_g$ of the~light wave are given by the refractive index $\eta$ and Eq. (\ref{eq:dispersion_plasma}):
\begin{align}
v_{ph}\equiv&\frac{\omega_L}{k_L}=\frac{c}{\eta}, & v_g\equiv&\frac{\partial \omega_L}{\partial k_L}=\eta c,  & \text{where }\eta=&\sqrt{1-\frac{\omega_p^2}{\omega_L^2}}<1.
\label{eq:index_of_refrac_plasma}
\end{align}
The phase velocity of the wave in the plasma is set by its driver, i.e., by the group velocity of the laser pulse. The associated relativistic gamma factor of the wave $\gamma_p$ is accordingly given by
\begin{equation}
\gamma_p=\frac{1}{\sqrt{1-\left(\frac{v_g}{c}\right)^2}}\approx\frac{\omega_L}{\omega_p}.
\label{eq:gamma_factor_wave}
\end{equation}

\subsection{Modulation of Intense Light Pulses in Plasma}
\label{ch:LaserModulation in plasma}

At high intensities ($a_0 \gtrsim 1$) the laser modifies the plasma density (and electron mass) via the ponderomotive force. This in turn changes the local refractive index and leads to a phase modulation of the pulse, giving rise to intensity-dependent propagation effects.
\subsubsection{Relativistic Self-Focusing and Guiding}
In an underdense plasma, the electrons are driven to relativistic energies by the strong laser fields. This modifies the plasma frequency via the relativistic mass increase as well as via the density perturbation by the ponderomotive force: 
\begin{equation}
\omega_{p,rel}^2\propto\frac{\omega_p^2}{\gamma}\frac{n_e}{n_{e,0}}.
\end{equation} 
According to Eq. (\ref{eq:index_of_refrac_plasma}) this change also modifies the index of refraction $\eta$. In general, as the trajectory of the electrons is mainly determined by their quiver motion ($\gamma=\sqrt{1+a^2}$), the refractive index can be approximated as \cite{bib:Esarey1996}
\begin{equation}
\eta(r,z)
\approx 1-\frac{1}{2}\left(\frac{\omega_p}{\omega_L}\right)^2\left(1-\frac{ a^2(r)}{2}+\frac{\delta n_e(r)}{n_{e,0}}
+\frac{\Delta n_{ext}(r)}{n_{e,0}}
\right).
\label{eq:refractive_index_long}
\end{equation}
Here, $\delta n_e(r)/n_{e,0}$ refers to the radial density perturbation by the ponderomotive force and leads to \textit{ponderomotive self-focusing}. Similarly, $\Delta n_{ext}(r)/n_{e,0}$ accounts for a pre-formed plasma channel, as can be formed by an axial discharge or an auxiliary laser pulse. 
 
The first term, $a^2(r)/2$ describes the influence of the electron gamma factor $\gamma_\perp$ on $\eta$. For a~laser intensity profile peaked on axis, $\partial a^2/\partial^2 r <0$, it leads to a reduced index of refraction on axis $\partial \eta/\partial r<0$ causing   \textit{relativistic self-focusing} (SF). 
This term counteracts the natural diffraction; thus, the laser pulse can be guided and sustain high intensities over several Rayleigh lengths. The threshold for the relativistic self-focusing can be found by balancing self-focusing with diffraction, leading to the condition $(a_0 k_p w_0)^2\geq32$, which can be associated to a critical power\footnote{keeping in mind that $P_L=\frac{\pi w_0}{2} I_L=\frac{\pi w_0}{2}\frac{\pi^2 \epsilon_0 m_e^2 c^5a_0^2}{e^2\lambda_L^2}$} $P_{c}$ \cite{bib:Mori1997, bib:Sun1987}
\begin{align}
P_{c}=&\frac{8\pi\epsilon_0m_{e}^2c^5\omega_L^2}{e^2\omega_p^2}=17.4\frac{\omega_L^2}{\omega_p^2}[\si{GW}]. 
\label{eq:crit_power}
\end{align} 
In presence of SF, the spatial evolution of a Gaussian pulse as given in Eq. (\ref{eq:waist_evolution}) has to be modified \cite{bib:Esarey2009}
\begin{equation*}
w(z)=\sqrt{1+\left(1-\frac{P}{P_c}\right)\frac{z^2}{z_R^2}}.
\end{equation*}
A laser pulse with a power $P$ focused to a spot size $w$ in a plasma diffracts for $P<P_{c}$, remains guided for $P=P_{c}$ and is focused for $P>P_{c}$. A `catastrophic' focusing is prevented by higher order effects \cite{bib:Hafizi2000,bib:Sprangle1987}. Such self-guiding effect has been experimentally observed in plasmas over several Rayleigh lengths \cite{bib:Ralph2009}. It is most effective when the spot size matches the blow-out radius, cf. Section~\ref{ch:bubble_regime}. For typical experimental parameters ($n_{e,0} = \SI{5e18}{\per\cubic\cm}$), the critical power is $P_{c}\simeq\SI{6}{\tera\watt}$, which is significantly smaller than the power provided by current laser systems. \\

\subsubsection{Temporal Pulse Modification}
The changes of the refractive index in propagation direction $\vec{e}_z$ can be written in a similar form as its transverse variation. With the co-moving coordinate $\xi=z-ct$ the refractive index is given by \cite{bib:Mori1997}
\begin{equation*}
\eta(\xi)
\approx 1-\frac{1}{2}\left(\frac{\omega_p}{\omega_L}\right)^2\left(1-\frac{a^2(\xi)}{2}+\frac{\delta n_e(\xi)}{n_{e,0}}-2\frac{\delta \omega(\xi)}{\omega_L}\right),
\end{equation*}
where the first term is identified with the longitudinal variation of the laser intensity. The last term $\delta\omega/\omega_L$ accounts for the change in the laser frequency across the laser pulse and describes the evolution of the~pulse chirp. The modulation in frequency follows the local change in the refractive index via \cite{bib:Mori1997}
\begin{equation*}
\frac{1}{\omega_L}\frac{\partial \omega}{\partial \tau}=-\frac{1}{\eta^2}\frac{\partial \eta}{\partial \xi}.
\end{equation*}
For Gaussian pulses, the gradients at the leading (falling) edge of the laser pulse will be red (blue) shifted causing a broadening of the spectrum, which is known as self-phase modulation (SPM). The~density change $\delta n_e(\xi)$ results in a varying local index of refraction. For a positive density slope, as encountered by the pulse in a typical LWFA experiment, the group velocity at the front of the pulse experiences a~deceleration, while the tail sees an acceleration. Such pulse shortening effects have been observed in experiments \cite{bib:Faure2005, bib:Schreiber2010} and can be approximated as
\begin{equation}
\tau_{comp}=\tau_0-\frac{n_{e,0}l}{2cn_{e,c}}, \label{eq:laser_duration_comp}
\end{equation}
where $l$ is the propagation length in the plasma and $\tau_0$ and $\tau_{comp}$ are the pulse durations before and after the interaction with the plasma. The variation of $\eta$ also modifies the group and phase velocity (\ref{eq:index_of_refrac_plasma}).
\begin{align*}
v_{ph}&\!=\!\frac{c}{\eta} \!\approx\! c\left[ 1\!+\!\frac{\omega_p^2}{2\omega_L^2}\left(1\!-\!\frac{\langle a^2\rangle}{2}\!+\!\frac{\delta n_e}{n_{e,0}}\!-\!\frac{2\delta \omega}{\omega_L}\right)\right], \;\;\; & 
v_g&\!=\!c\eta\!\approx\! c\left[ 1\!-\!\frac{\omega_p^2}{2\omega_L^2}\left(1\!-\!\frac{\langle a^2\rangle}{2}\!+\!\frac{\delta n_e}{n_{e,0}}\!-\!\frac{2\delta \omega}{\omega_L}\right)\right].
\end{align*} 
The combined effect of SF and SPM is a focusing and compression of the pulse, boosting its intensity. Since the ponderomotive potential causes a repulsion of electrons from regions of high intensity, the net effect looks as it the electron inertia would "squeeze" the laser into a smaller, more intense bullet of light. 

\subsection{Excitation of Plasma Waves}

A laser pulse excites a plasma wave by displacing electrons from their equilibrium position. 
In the~Coulomb gauge ($\nabla A=0$), the Lorentz Eq. (\ref{eq:Lorentzplasma}) in terms of the electromagnetic potentials reads:
\begin{equation}
\left( \frac{\partial}{\partial t}+ \vec{v} \cdot \nabla \right) \vec{p}=e\left[\frac{\partial}{\partial t}\vec{A}+\nabla \Phi - \vec{v} \times \nabla \times \vec{A}\right]. \label{eq:fluid}
\end{equation}
In addition, using the vector identity $\nabla p^2=2 \left[(\vec{p}\cdot \nabla)\vec{p}+\vec{p}\times\left(\nabla \times \vec{p}\right)\right]$ and  $\nabla p^2=m_e^2c^22\gamma\nabla\gamma$ (obtained from Eq. (\ref{eq:gammafactor})) yields 
\begin{equation*}
m_e c^2\nabla\gamma=(\vec{v}\cdot\nabla)\vec{p}+\vec{v}\times(\nabla\times\vec{p}). 
\end{equation*}
The term on the left represents the driving term, i.e., the ponderomotive force as introduced in Eq.~(\ref{eq:pond_force}).
Inserting the above expression into Eq.~(\ref{eq:fluid}), the equation of motion can be expressed as 
\begin{equation}
\boxed{
\frac{\partial \vec{p}}{\partial t}=e \nabla\Phi+e \frac{\partial \vec{A}}{\partial t}-m_ec^2\nabla\gamma \label{eq:eqofmotion}}
\end{equation}
Here, $\vec{v}\times\nabla\times(\vec{p}-e\vec{A})$ is set to 0, indicating that the electrons are only driven by the vector potential. This can be understood by realizing that the curl of Eq.~(\ref{eq:fluid}) leads to $\partial (\vec{p}-e\vec{A})/\partial t=\vec{v}\times(\nabla\times(\vec{p}-e\vec{A}))$. Therefore, the canonical momentum $\vec{p}-e \vec{A}=0$ remains zero for all times if there is no initial perturbation of the plasma and the electrons initially have zero canonical momentum ($\vec{p_0}-e\vec{A_0}=0$) \cite{bib:Chen1993}. This equation states the starting point for exploring the solution of the linearly and nonlinearly driven plasma waves.
Moreover, for the upcoming derivations it is convenient to use normalized quantities 
\begin{align}
\beta&=\vec{v}/c, & \vec{a_0}&=\frac{e\vec{A}}{m_ec}, & \phi&=\frac{e\Phi}{m_e c^2}, & \gamma&=\frac{E}{m_ec^2}, & \vec{u}&=\frac{\vec{p}}{m_ec}. 
\end{align}
and to introduce a coordinate transformation to a frame moving with the laser pulse at a speed $v_g$. The~new variables $(\xi,\tau)$ are given by $\xi=z-v_gt$ and $\tau=t$ and their partial derivatives are:
\begin{align}
\frac{\partial}{\partial z}&=\frac{\partial}{\partial \xi} &\text{and}& &\frac{\partial }{\partial t}&=\frac{\partial }{\partial \tau}=-v_g \frac{\partial}{\partial \xi}\overset{v_g\rightarrow c}{\approx} -c\frac{\partial}{d\xi}.
\label{eq:partial_derivative}
\end{align}

\subsubsection{Linear Regime} 

An analytical solution for small laser intensities ($a_0 \ll 1$) and weakly perturbed plasmas ($\delta n_e \ll n_{e,0}$) can be derived from the continuity and Poisson Eqs. (\ref{eq:cont.eq}), (\ref{eq:poisson}). The solution for the scalar potential $\phi$, yields for a Gaussian laser envelope $a=a_0\exp(-\xi^2/(c\tau_0)^2)\exp(-r^2/w_0^2)$ after the laser transit
\begin{align}
\phi(r,\xi)\!=\!-f(r)\sin(k_p\xi), \;\;\;\;\;\;\;\; \;\;\;\;\;\;\;\; f(r)\!=\!a_0^2\sqrt{\frac{\pi}{2}}\frac{k_p}{4} c\tau_0\exp\left(-\frac{2r^2}{w_0^2}\right)\exp\left(-\frac{(k_p c \tau_0)^2}{8}\right).
\label{eq:linear_phi}
\end{align}
From the definition of the scalar potential $\phi$, Eq. ~(\ref{eq:vectorpot}) and Poisson's Eq. (\ref{eq:poisson}), the electric field and the~electron density are given as:
\begin{align}
\frac{E_z}{E_{p,0}}=&-\frac{1}{k_p}\frac{\partial \phi}{\partial \xi}, &
\frac{E_r}{E_{p,0}}=&-\frac{1}{k_p}\frac{\partial \phi}{\partial r}, &
\label{eq:linear_Ez}
\frac{\delta n_e}{n_{e,0}}=&\frac{1}{k_p^2}\frac{\partial^2 \phi}{\partial \xi^2}.
\end{align}
$E_{p,0}$ denotes the cold nonrelativistic wavebreaking field (see Eq. (\ref{eq:wave_breaking_field_nonrel})).
\begin{figure}
    \subfigure[\textbf{Top:} Normalized plasma potential $\phi$, longitudinal electric field $E_z/E_0$ and density perturbation $\delta n_e/n_{e,0}$ on axis ($r=0$). \textbf{Bottom:} plasma density map $\delta n_e(r,\xi)/n_{e,0}$ generated by the ponderomotive force of a Gaussian laser focus.]
    {\includegraphics[width=0.47\textwidth]{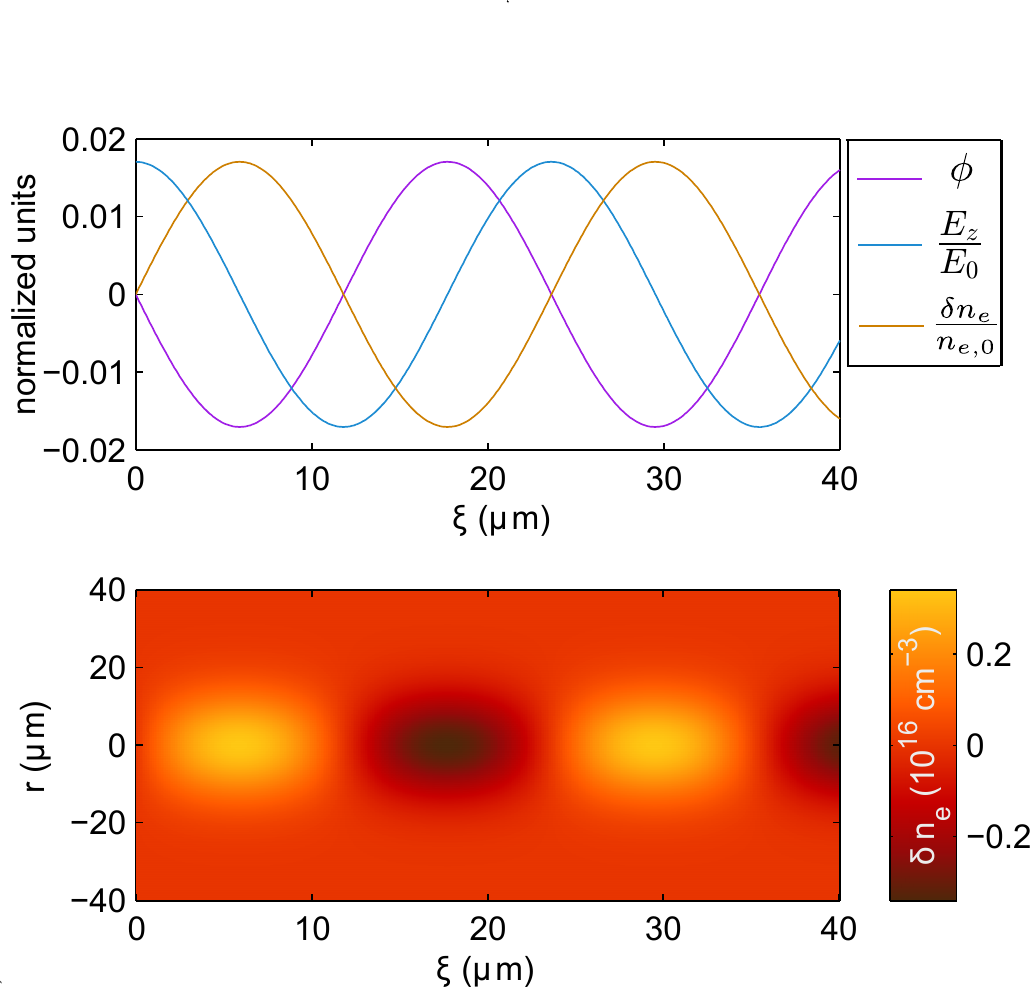}
        \label{fig:linearsolution_density}
        }
    \hspace*{0.02\textwidth}
    \subfigure[Spatial extend of the longitudinal $E_z(r,\xi)$ (on the~\textbf{Top}) and the radial electric field $E_r(r,\xi)$ (on the \textbf{Bottom}). The~green area marks a $\lambda_p/4$-phase region of the wakefield with an accelerating and transverse focusing field.]
    {\includegraphics[width=0.47\textwidth]{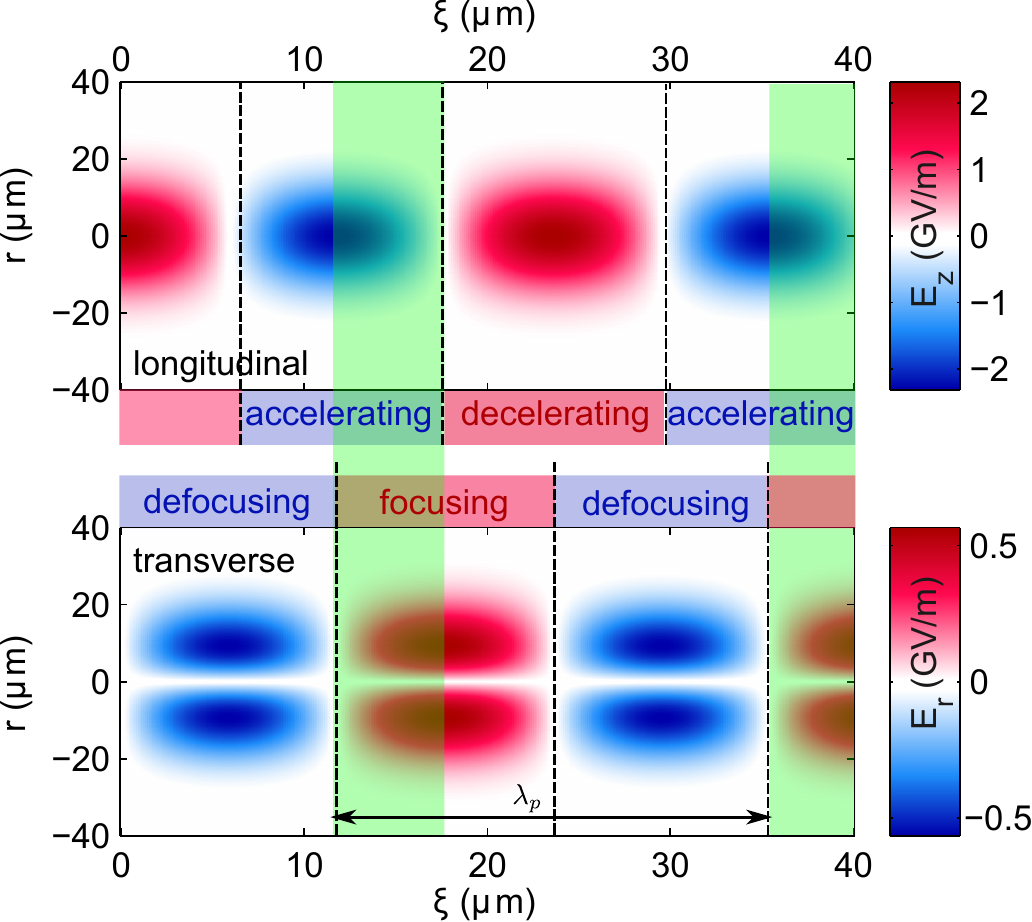}
        \label{fig:linearsolution_fields}
    }
      \caption[\textbf{3D Linear wakefield quantities} \newline
    evaluation :	\textbackslash Th LaserPlasma\textbackslash Linear 3D Solution.m   \newline
    figure file a:		\textbackslash Th LaserPlasma\textbackslash 3D density final.pdf \newline
    figure file b:	\textbackslash Th LaserPlasma\textbackslash 3D Electric field final.pdf     ]  
    {\textbf{3D linear wakefield quantities in the moving co-ordinate system} created in the focus of a laser pulse ($a_0=0.2$, $t_{FWHM}=\SI{28}{fs}$ and $d_{FWHM}=\SI{22}{\micro\meter}$) for a plasma density of $\SI{2e18}{\per\cubic\cm}$.}
    \label{fig:linearsolution}
\end{figure}
\noindent
The solution for the transverse and longitudinal electric fields as well as the density perturbation excited by a weak Gaussian laser pulse ($a_0=0.2$) are displayed in Fig. \ref{fig:linearsolution}. The response of the normalized plasma quantities is linear (Fig. \ref{fig:linearsolution_density}). The wakefield radius is determined by the focal spot size.
Figure 2(b) denotes alternating phase regions where the wake is either accelerating/decelerating or focusing/defocusing.  LWFA takes place where focusing and accelerating phases coincide, i.e. in the shaded phase intervals with a width $\lambda_p/4$. 
The peak wakefield potential is determined by the laser intensity and pulse duration: $\phi \propto a_0^2\tau_0$. For a given electron density $n_{e,0}$, the wave is driven resonantly ($\phi\rightarrow\phi_{max}$) for a pulse duration of 
$t_{FWHM}^{opt}[\si{fs}]\approx 41.2(n_{e,0}[\SI{e18}{\per\cubic\cm}])^{-\sfrac{1}{2}}$, corresponding to a laser pulse length of $L_{FWHM}=0.37\lambda_p$. This requires short laser pulses for electron acceleration at a plasma wavelength of $\lambda_p[\si{\micro\meter}]=33.4(n_{e,0}[\SI{e18}{\per\cubic\cm}])^{-\sfrac{1}{2}}$. The accelerating and focusing fields in this weakly relativistic regime reach gradients of \si{\giga\electronvolt\per\meter} at $a_0 = 0.1$, and further increase with laser intensity. For $a_0\rightarrow1$ the wave cannot be treated by linear perturbation theory anymore. 

\subsubsection{Non-linear Regime}
At high intensities ($a_0>1$), the plasma response becomes highly nonlinear and has to be treated non-perturbatively. The general case of arbitrary pump strengths has no 3D-solution, and reduced models only have limited predictive power. Therefore,  phenomenological scaling laws based on 3D particle-in-cell (PIC) simulations \cite{bib:Gordienko2005,bib:Lu2007, bib:Ding2020}, see also Section \ref{ch:scaling_laws}, are  employed.
However, analytical solutions exist in the 1-D spatial case \cite{bib:Esarey1997,bib:Sprangle1990}, that can explain certain aspects of nonlinear plasma waves and help understanding their properties. In 1-D space,  the 3-D electron momentum can be decomposed into a longitudinal $z$-component and a transverse component in the ($x$,$y$)-plane. The equation of motion in both directions can be derived \cite{bib:Gibbon2005} from Eqs. (\ref{eq:Lorentz}) and (\ref{eq:energy_equation}):
\begin{align*}
\frac{d\vec{p_\perp}}{d t}&=e(\vec{E_\perp}+\vec{v_z} \times \vec{B_z})=e \frac{d \vec{A_\perp}}{d t},  & \frac{dE}{dt}=c\frac{dp_\parallel}{dt}.
\end{align*}
Integration for an initially resting electron yields for the transverse and longitudinal momentum
\begin{align}
\vec{ p_\perp}=e\vec{A_\perp}\; \Leftrightarrow\; \vec{u_\perp}=\gamma \vec{\beta_\perp}=\vec{a}, \label{eq:vectorpotential_perp}\\
E-cp_\parallel=m_ec^2\; \Leftrightarrow\; \gamma-1=u_\parallel.
\label{eq:vectorpotential_parallel}
\end{align}
As in the linear case, the solution for the wake potential $\phi$ for an arbitrary pump can be deduced from Eq.~(\ref{eq:eqofmotion}), the continuity Eq.~(\ref{eq:cont.eq}) and Poisson's Eq.~(\ref{eq:poisson}) 
\begin{equation}
\frac{\partial^2 \phi}{\partial \xi^2}=k_p^2\gamma_p^2\left(\beta_p\left(1-\frac{1+a^2}{\gamma_p^2\left(1+\phi\right)^2}\right)^{-\sfrac{1}{2}}-1\right) .\label{eq:poissonincomov}
\end{equation}
The above nonlinear ordinary differential equation can now be solved for the potential $\phi$. The electric field and density perturbation are again obtained from Eq.~(\ref{eq:linear_Ez}).
For a linearly polarized  square pulse with $v_g \rightarrow c$ the scalar potential $\phi_{0,max}$ and the peak electric field $E_{0,max}$ are given by\cite{bib:Berezhiani1990}:
\begin{align}
\phi_{0,max} \approx a_0^2/2, & & E_{0,max} = E_{p,0} \frac{a_0^2/2}{\sqrt{1+a_0^2/2}}.
\label{eq:max_electric_field}
\end{align}
For an arbitrary pulse envelope $a(\xi)$ the potential, the field and the density are obtained numerically. The~solution for a Gaussian laser pulse in the linear ($a_0 \approx 0.4$) and nonlinear ($a_0 \approx 2.1$) regime are given in the top panels of Fig. \ref{fig:separatrix_nonlinear}. While nearly sinusoidal in the linear case, the non-linear density distribution exhibits narrow peaks with a periodicity of $\lambda_{p,rel}>\lambda_p$, separated by low-density regions. The electric field has a sawtooth shape and is linear over most of the period. The non-linear plasma wavelength $\lambda_{p,rel}$ inceases with $a_{0}$. For a squared pulse and 
$\gamma_p\gg1$, 
the solution for $\lambda_p$ can be found analytically \cite{bib:Esarey2009}
\begin{align}
\lambda_{p,rel}= \lambda_p \begin{cases} 1+\frac{3}{16}\frac{a_0^2}{\sqrt{1+a_0^2}}  &\text{for } a_0^2 \ll 1\\
\frac{2}{\pi}\left[\frac{a_0^2}{\sqrt{1+a_0^2}} +\frac{\sqrt{1+a_0^2}}{a_0^2}\right]& \text{for } a_0^2 \gg 1\\ \end{cases}.
\label{eq:nonlinear_plasmawavelength}
\end{align}

\subsection{Wavebreaking}

The strength of the accelerating field defines the maximum energy gain for a fixed acceleration length. It depends on the density perturbation $\delta n_e/n_{e,0}$ of the plasma electrons in the density peaks. With increasing  driver strength, the speed of the background electrons can exceed the phase velocity of the~plasma wave, leading to a breakdown of the wakefield structure called \textit{longitudinal} wavebreaking. The  corresponding maximum field $E_{wb}$ limits the validity of the fluid plasma model.

In the linear regime, $E_{wb}$ can be estimated by assuming that all electrons of the wave ($\delta n_e=n_{e,0}$) oscillate with $\omega_p$. Solving Eq.~(\ref{eq:Max1}), $\nabla_\parallel E\approx-\frac{e}{\epsilon_0}n_{e,0}$ yields the \textit{cold, nonrelativistic wavebreaking field} \cite{bib:Dawson1959}
\begin{align}
E_{wb}=E_{p,0}=&\frac{m_ec\omega_p}{e}, & E_{p,0}[\si{\giga\volt\per\m}]=96\sqrt{n_{e,0}[\SI{e18}{\per\cubic\cm}]}.
\label{eq:wave_breaking_field_nonrel}
\end{align}
For relativistic fluid velocities nonlinear plasma waves can easily exceed $E_{wb}$. The analytic solution can be found if the electron density becomes singular in Eq.~(\ref{eq:poissonincomov})  ($\frac{\partial^2 \Phi}{\partial^2 \xi} \rightarrow \infty $), i.e., $\gamma_\perp =\sqrt{1+a^2}= \gamma _p\left(1+\phi\right)$. On the other hand the minimum/maximum potential for the partial differential equation (PDE) behind the driving laser ($\gamma_\perp=1$) is related to the maximum electric field $E_{z,max}$ via \cite{bib:Schroeder2006}
\begin{equation}
\phi_{min/max}=\frac{1}{2}\left(\frac{E_{z,max}}{E_{p,0}}\right)^2\pm\beta_p\left[\left(1+\frac{1}{2}\left(\frac{E_{z,max}}{E_{p,0}}\right)^2\right]^2-1\right]^{\sfrac{1}{2}}.
\label{eq:separatrix_phi}
\end{equation}
Equalizing $\phi_{min}$ and $\phi=1/\gamma_p-1$ yields the \textit{cold relativistic wake-breaking field} \cite{bib:Esarey1995}
\begin{equation}
E_{wb,rel}=E_{p,0}\sqrt{2(\gamma_p-1)}.
\label{eq:wave_breaking_field_rel}
\end{equation}
Here, as in the nonrelativistic case, the background electrons can outrun the wake's phase velocity and are injected. This \textit{self-injection}  generally continues once it has started, causing a large energy spread. In three dimensions, the wave also breaks transversely, which generally happens at much lower $a_0$ than in this purely 1-D treatment. For high beam quality,  it is sometimes desired to avoid self-trapping by operating below the wavebreaking threshold, and relying on other injection methods (see Section \ref{ch:Electron_injection}).

\section{Laser Wakefield Acceleration}
The fundamental role of the plasma in LWFA is to transform the strong, transversely oscillating laser fields into quasistatic, longitudinally accelerating fields. The longitudinal dynamics of the electrons, i.e. their orbit in phase space, $\left(u_z,\xi\right)$, by be treated by the Hamiltonian formalism.
\subsection{Electron Acceleration}
\label{ch:Electron_Acceleration}
\begin{figure}
    \subfigure[Electron orbits in the nonrelativistic regime ($a_0<1$) feature a nearly symmetric separatrix with a width of $\simeq \lambda_p$. The maximum energy gain is $\Delta u_z \approx 2\gamma_p^2 \Delta \phi=1.6\times10^2$.]
    {\includegraphics[width=0.47\textwidth]{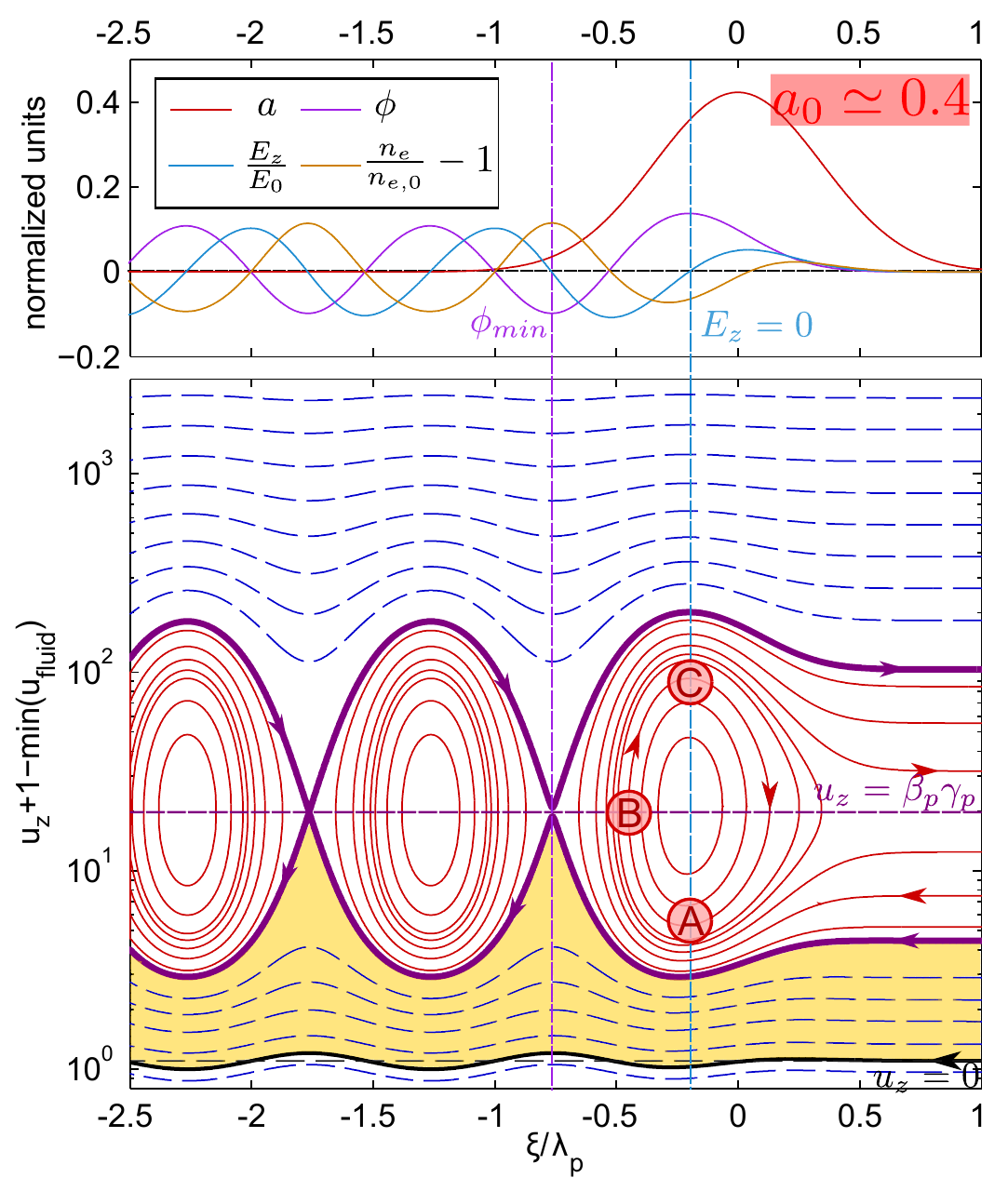}
        \label{fig:separatrix_linear}
        }
    \hspace*{0.02\textwidth}
    \subfigure[Electron orbits in the relativistic regime with an asymmetric separatrix. Its width is given by the nonlinear plasma wavelength $\lambda_{p,rel}\sim 1.3\lambda_p$. The maximum energy gain is $\Delta u_z\approx2.25\times10^3$.]
    {\includegraphics[width=0.47\textwidth]{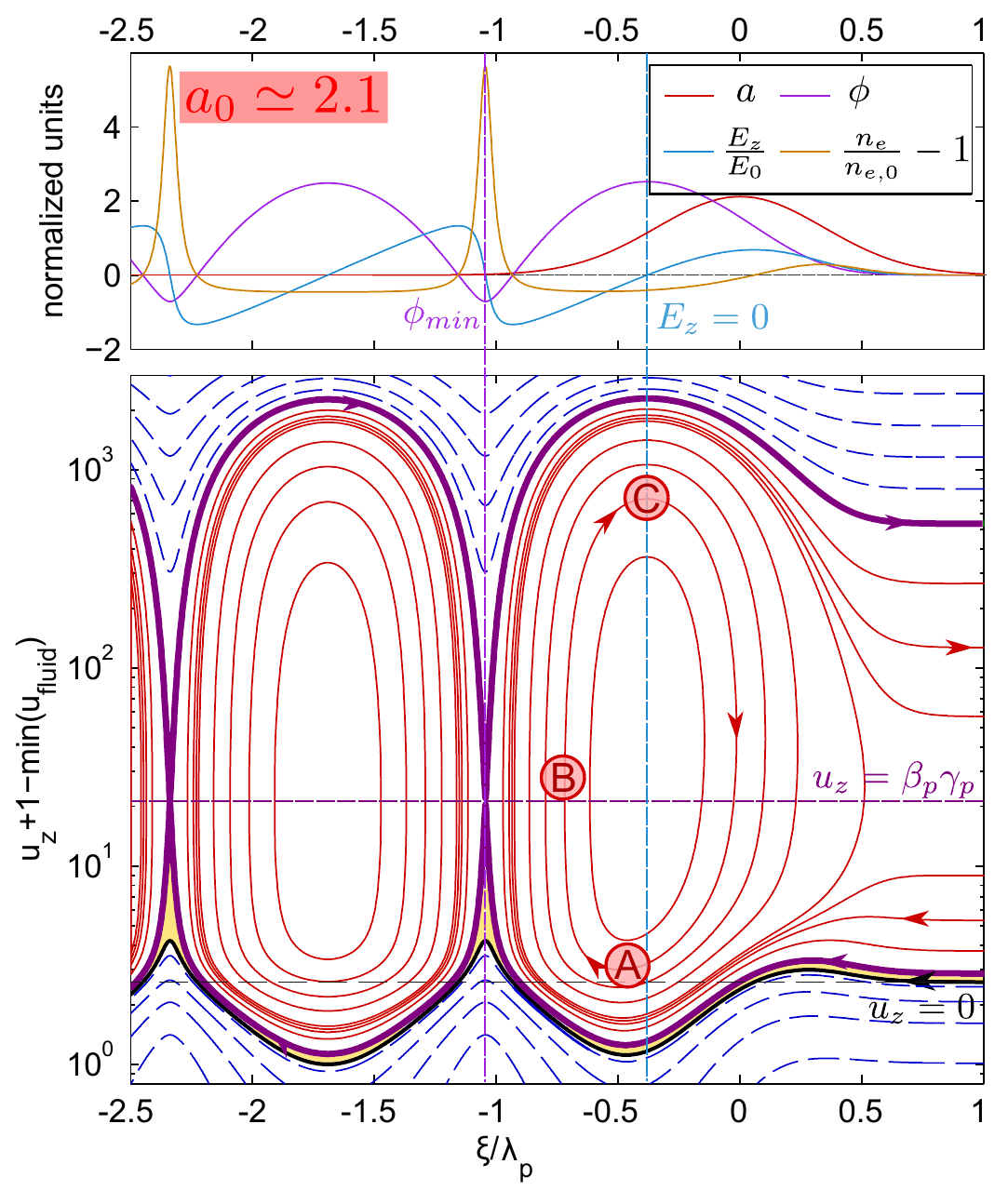}
        \label{fig:separatrix_nonlinear}
    }
    \caption
    [\textbf{Linear and nonlinear separatrix} \newline
    evaluation :	\textbackslash Th LaserPlasma\textbackslash Separatrix.m   \newline
    figure file a:		\textbackslash Th LaserPlasma\textbackslash Separatrix lin final.pdf/svg \newline
    figure file b:	\textbackslash Th LaserPlasma\textbackslash Separatrix nonlin final.pdf/svg      ]  
    {\textbf{Longitudinal electron phase space trajectories in a laser-driven plasma  wave } with $t_{FWHM}=\SI{28}{fs}$, $a_0=0.6/\sqrt{2}$ (\textbf{left}) and $a_0=3/\sqrt{2}$ (\textbf{right}). Trapped orbits (red solid lines), the separatrix (purple solid line), fluid (solid black line) and untrapped electron orbits (dashed blue lines). The corresponding  normalized wakefield quantities are plotted at the top of each figure. The shaded area marks the momentum gap to be overcome by fluid electrons to get trapped. The background electron density in both plots is $n_{e,0}=\SI{5e18}{\per\cubic\cm}$, i.e., $\gamma_p=19$.}
\label{fig:separatrix}
\end{figure}

The 1D Hamiltonian for an electron interaction with a laser field in a plasma wave in its normalized quantities is given by \cite{bib:Esarey1995, bib:Schroeder2006}
\begin{equation*}
\mathcal{H'}\left(z,u_z \right)=\sqrt{1+u_\perp^2+u_z^2}-\phi(z-v_gt).
\end{equation*}
The first root term describes the normalized kinetic energy $\gamma$. It can be further simplified by observing Eq. (\ref{eq:vectorpotential_perp}). The second term represents the potential energy. Here, the time dependence can be eliminated by a canonical transformation of the Hamiltonian $(z,u_z) \rightarrow(\xi,u_z)$\footnote{Using a generating function $F(z,u_z)=u_z\times(z-v_gt)$ the new Hamiltonian reads $\mathcal{H}=\mathcal{H'}-\frac{1}{c}\frac{\partial F}{\partial t}$.}, giving:
\begin{equation}
\mathcal{H}\left(\xi,u_z \right)=\sqrt{1+a^2\left(\xi\right)+{u_z(\xi)}^2}+\phi\left(\xi\right)-\beta_p {u_z(\xi)}.
\label{eq:Hamiltonian}
\end{equation}
$\mathcal{H}\left( \xi,u_z\right)=H_0=const.$ describes the motion of an electron on a distinct orbit in the plasma wave. Solving Eq. (\ref{eq:Hamiltonian}) for $u_z(\xi)$ gives the trajectory of the electron in the phase space
\begin{equation}
{u_z}=\beta_p\gamma_p^2\left(H_0+\phi\right)\pm\gamma_p\sqrt{\gamma_p^2\left(H_0+\phi\right)^2 -\gamma_\perp^2}.
\label{eq:separatrix_orbit}
\end{equation}
$u_z$ represents an electron orbit of constant total energy for a given set of $a(\xi)$, $\phi(\xi)$ and $H_0$.  Examples are plotted in the lower part of Fig. \ref{fig:separatrix}. The lines represent the possible electron trajectories ($\xi,u_z$) with varying $H_0$ in a linear ($a_0<1$) and non-linear ($a_0>1$) regime. 
In general, two types of orbits can be distinguished: the closed orbits (solid red) describe the trapped and the open orbits (dashed blue) the untrapped electrons. The separatrix, displayed by the thick purple line, denotes the trajectory where open and closed orbits meet, i.e., where the radicand of Eq. (\ref{eq:separatrix_orbit}) is equal to zero. It occurs at $\phi(\xi_{min})=\phi_{min}$, indicated by a thin vertical line. Therefore, the Hamiltonian of the separatrix is $H_{sep}=\gamma_\perp(\xi_{min})/\gamma_p-\phi_{min}$. 
The~absolute minimum $u_z$ can be derived from Eq. (\ref{eq:separatrix_orbit}) to $u_z(\xi_{min})=\beta_p\gamma_p$.

Electrons initially at rest, $u_\perp(\xi=+ \infty)=u_z(\xi=+ \infty)=0$, are characterized by the Hamiltonian $H_{fluid}=1$. These fluid electrons,  (thick black line) constitute the plasma wave and are pushed by the~ponderomotive force without gaining any substantial momentum from the wave. Similarly, the~electrons with a too low/high initial momentum (dashed blue lines) are moving on open orbits and are not trapped by the wake. The latter follow the condition $H_0>H_{sep}$. 

In contrast, an electron with a sufficient initial momentum $u_\perp(\xi =+\infty)>u_{z,min}$ and $H_0<H_{sep}$, (drawn in red), is located on closed orbits. It is trapped in the wave and successively gains and looses energy while performing a rotation in phase space.  At {\larger\textcircled{\smaller[2]A}}, the electron starts gaining energy, while it slips back in longitudinal position ($\beta_e<\beta_p$) until {\larger\textcircled{\smaller[2]B}}. Here, the electron has reached the plasma wave velocity and starts to outrun the wake ($\beta_e>\beta_p$). At {\larger\textcircled{\smaller[2]C}} it has maximum energy and, unless the~acceleration process is terminated, enters the region with decelerating fields ($E_z>0$). This process is called dephasing and sets one of the limitations for the electron energy gain in LWFA, cf. Section  \ref{ch:limitsofLWFA}.\\
The highest momentum gain is experienced by electrons following the separatrix orbit, i.e.,
\begin{equation*}
\Delta u_z=u_{z,max}-u_{z,min}\approx 2\gamma_p^2 \Delta \phi.
\end{equation*}
Thus, the energy gain of the electrons is dependent on the electron density and the laser intensity. As long as the electrons are injected and extracted at the right phase the acquired energy can exceed the~$\SI{}{\giga\electronvolt}$-barrier, cf. Fig. \ref{fig:separatrix_nonlinear}.

\subsection{Electron Injection}
\label{ch:Electron_injection}
As indicated above, the details of the injection process largely define the overall beam quality, i.e., the~energy spread, charge, and divergence.
In order to achieve minimal emittance, the electrons should be injected with sufficient initial momentum into a spatial volume that is matched to the beam charge and background density. This is difficult to achieve with an external (RF) electron source, because it requires matching a beam from the RF injector with structure scales of centimeters to the plasma accelerator with 10's of $\SI{}{\micro\meter}$.  External injection has been demonstrated into low-density plasmas with long wavelength\cite{bib:Clayton1993, bib:Amiranoff1998}, but with currect high-intensity Ti:Sa lasers the comparably high densities so far have only allowed successful injection very recently \cite{bib:Hua2019}.

From an experimentalist's point of view, \textit{self-injection} appears to be the most straighforward approach. It is based on wave-breaking, sometimes augmented by the expansion of the cavity size during self-focusing of the drive laser. This slows down the phase velocity of the first wakefield peak and allows electrons to be trapped more easily over a considerable distance, leading to electron beams with a large momentum spread.
Therefore, a localized and controlled injection process is key for achieving narrow-band electron bunches. Many schemes have been studied and still are an active area of research. Three of them (in addition to self-injection) will be described in the following sections:
\begin{itemize}
\item \textit{Density transition injection}. A localized electron density down-ramp  increases $\lambda_p$ and reduces the~wake's phase velocity, causing a defined injection.
\item \textit{Colliding pulse injection}. A locally confined beat wave of two colliding laser pulses gives background electrons a momentum kick and injects them into the proper phase.
\item \textit{Ionization injection}. Electrons from core levels of heavy atoms are ionized and freed close to the~peak of the laser pulse; thus, they are `born' in a trapped orbit.
\end{itemize}

\subsubsection{Self-Injection}
\label{ch:self_injection}
In a purely 1-D model, the minimum initial energy for electrons to be trapped is given by
\begin{align}
W_{trap}=m_ec^2\left(\sqrt{1+u_{z,sep}^2(+\infty)}-1\right),
\label{eq:trapping_energy}
\end{align} 
where $u_{z,sep}(+\infty)$ describes the separatrix orbit in front of the laser pulse ($ a_0=\phi=\vec{u_\perp}=0$)
\begin{equation*}
u_{z,sep}(+\infty)=\beta_p\gamma_p^2H_{sep}-\gamma_p\sqrt{\gamma_p^2H_{sep}^2-1}.
\end{equation*}
Figure \ref{fig:injection_threshold} shows the dependence of the trapping energy $W_{trap}$ (Eq.~(\ref{eq:trapping_energy})) on the normalized maximum electric field $E_{z,max}/E_{p,0}$ with $E_{z,max}$ given by Eq.~(\ref{eq:separatrix_phi}). 
On the top $x$-axis, the $W_{trap}$-dependence with respect to the wave velocity $\gamma_p$ given by Eq.~(\ref{eq:gamma_factor_wave}) is displayed for various $\phi_{min}$. The threshold decreases for larger wakefield amplitudes and slower wake phase velocities, i.e. for high laser intensities and high electron densities.
With increasing $a_0$ the fluid orbits get closer to the separatrix, and the necessary momentum for trapping can be significantly reduced. When the fluid and the separatrix orbit overlap, i.e., the longitudinal velocity of the electrons overtakes the plasma wave velocity, the wavebreaking limit of the plasma is reached, and the background electrons are directly injected. For common electron densities of $5 \times 10^{18} cm^{-3}$ this occurs at rather high values of $a_0 \approx 6$. Within a 1-D description, this cold wavebreaking limit can be reduced by thermal effects \cite{bib:Esarey2007}, but still remains higher than observed in most experiments..\\
\subsubsection{Longitudinal Self-Injection}
The situation changes quite dramatically in a 3D-dimensional geometry. At moderate laser intensities ($a_0\sim1-3$), when the laser forms a wake without reaching the blow-out regime, (cf. Section \ref{ch:bubble_regime}). Background electrons close to the axis can now tunnel through the decelerating field and be injected without a substantial transverse motion \cite{bib:Corde2013b}. 
Upon its propagation through the medium, the laser self-focuses  and compresses while $a_0$ increases. This lengthens the first wakefield bucket and reduces the phase velocity \cite{bib:Kalmykov2009,bib:Kostyukov2010}, and electrons can catch up with the plasma wave more easily and are injected at much lower $a_0$. Since this happens at an early stage of the acceleration with a small transverse momentum, the generated bunches exhibit good shot-to-shot stability, high energies and few-mrad divergence. The~longitudinal injection mechanism is dominat at low plasma density, where self-focusing is weak and the pulse diffracts after the first injection event without causing further (transverse) self-injection.\\

\subsubsection{Transverse Self-Injection}
At high laser intensities and electron densities, transverse wavebreaking can occur \cite{bib:Bulanov1997, bib:Modena1995}. The strong ponderomotive force kicks electrons radially away from the axis, forming a bubble-shaped electron cavity whose focusing field pulls them back onto the axis, forming a density spike. Here, some electrons can be scattered forward into an accelerating phase \cite{bib:Kostyukov2010}. 
The additional Coulomb repulsion of these injected electrons at the rear of the bubble lengthens the cavity and causes further injection. 
In contrast to longitudinal self-injection, these injected electrons enter the wakefield with a large transverse momentum, resulting in high divergence. 
As the laser spot size (and peak $a_0$) usually oscillate around the  matched spot size of the plasma \cite{bib:Corde2013b}, transverse injection can happen several times during the laser pulse evolution at positions with the highest $a_0$ before the bubble eventually becomes beam loaded, cf. Section \ref{ch:limitsofLWFA}. Due to these oscillations injection happens at random locations, resulting in low reproducibility.

The existence of a density threshold between longitudinal and transverse self-injection allows switching on and off the latter \cite{bib:Corde2013b}. Below the threshold, the electrons are trapped longitudinally, with low charge and a stable, sometimes narrowband spectrum. Above, transverse injection dominates by far, generating less stable electron beams with high charge. According to \cite{bib:Mangles2012}, self-injection occurs if 
\begin{equation}
\alpha_L W_L>\frac{\pi\epsilon_0m_e^2c^5}{e}\left[\ln\left(\frac{2n_{e,c}}{3n_e}\right)-1\right]^3\frac{n_{e,c}}{n_e}\tau(l),
\end{equation}
where $\alpha_L$ is the fraction of the laser energy $W_L$ contained within the FWHM of the focal spot and $n_e$ is the plasma density. $\tau(l)$ is the self-compressed laser pulse duration as given in Eq.~(\ref{eq:laser_duration_comp}).

The main advantage of self-injection is its experimental simplicity. However, it often lacks stability, as it relies on nonlinear self-focusing and self-compression processes. The generated electron beams exhibit fluctuating spectra with rather large momentum spread ($30-100\%$).

\begin{figure}
    \subfigure[\textbf{ Injection threshold } plotted on a semi-logarithmic scale vs maximum electric field for different electron densities $n_{e,0}=(15,5,1)\times10^{18}\; \si{\per\cubic\cm}$, corresponding to $\gamma_p=11,19,42$  (straight lines), as well as versus $\gamma_p$ for different minimum normalized potentials $\phi_{min}$ (dash-dotted lines). The injection threshold is lowered for larger wakefields and slower phase velocities. The dashed lines ($\gamma_p=19$) give the trapping threshold for the two cases in Fig.\ref{fig:separatrix}, $E_{trap}^{a_0\simeq0.4}\approx\MeV{1.3}$ and $E_{trap}^{a_0\simeq2.1}\approx\keV{20}$. ]
    {\includegraphics[width=0.47\textwidth]{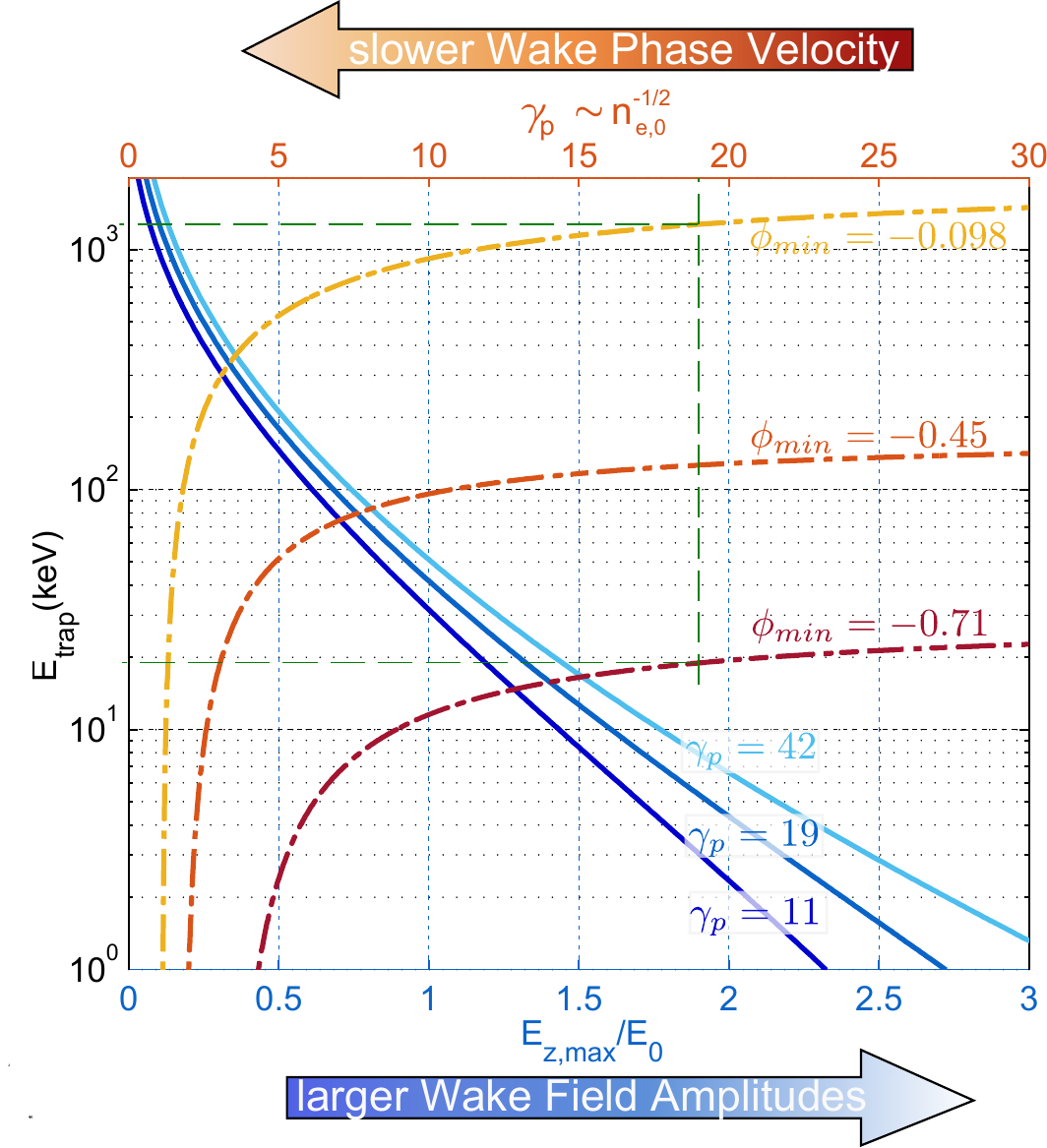}
        \label{fig:injection_threshold}
        }
    \hspace*{0.02\textwidth}
    \subfigure[\textbf{Shockfront injection mechanism}. The laser ($a_0=1.5$ $\SI{28}{fs}$) interacts with a density profile as shown on the top. In the high density region {\larger\textcircled{\smaller[2]\textcolor{DarkGreen}{1}}} with $\SI{5e18}{\per\cubic\cm}$ it drives a plasma wave with $\lambda_{p,1}$ and creates the phase orbits shown dashed in the bottom plot. After the transition the plasma wavelength lengthens to $\lambda_{p,2}$ (with lower $n_e=\SI{3.5e18}{\per\cubic\cm}$) in {\larger\textcircled{\smaller[2]\textcolor{DarkGreen}{2}}}. This sudden expansion causes the untrapped electrons in the wave peaks to rephase inside the new separatrix (solid purple line).]
    {\includegraphics[width=0.47\textwidth]{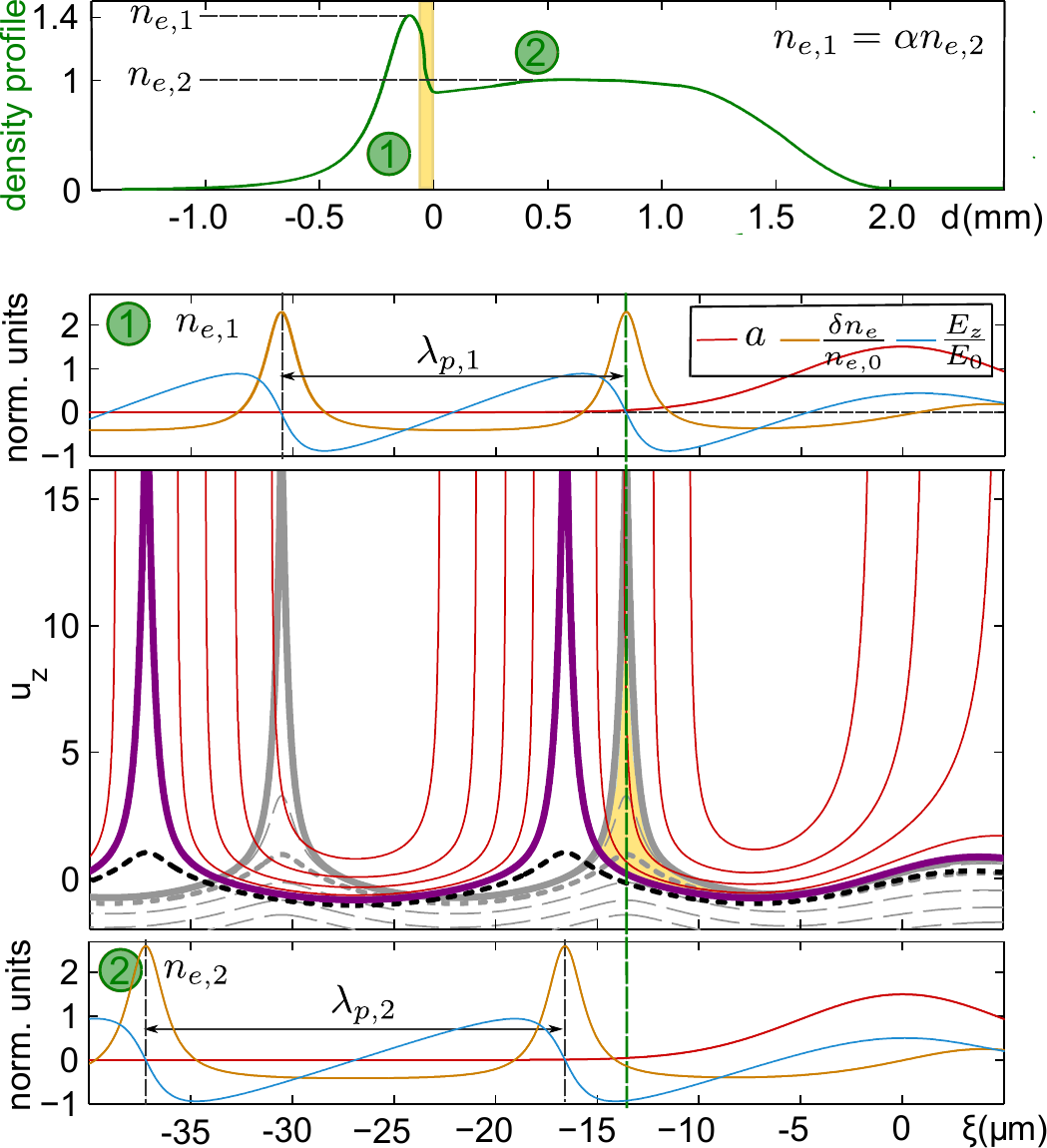}
        \label{fig:shockfront_injection}
    }
 
\caption { Injection threshold for the self-injection (\textbf{left}) and illustration of shockfront injection (\textbf{right}).}
    \label{fig:}
\end{figure}
\subsubsection{Injection in Density Transitions}
\label{ch:shock-front injection}
In contrast, reproducible low-emitance, low-energy spread beams require externally controlled  injection schemes, where in most cases the main LWFA is operated below the threshold for wavebreaking. The~injection process itself is triggered locally at a defined position along the propagation direction by tailoring the longitudinal density profile inside the~plasma. In negative density transitions ($\partial n_e/\partial z<0$) the~plasma wave elongates according to $\lambda_p \propto n_e^{\sfrac{-1}{2}}$ and the back of the~plasma wave slows down momentarily, thus lowering the injection threshold and causing self-trapping. After the down-gradient, the~laser drives a wake in low density which accelerates the injected electrons, but avoids self-trapping. The control over the density profile therefore allows to separately control injection and acceleration.

For a \textit{long density transition}, $L_{tr}>\lambda_p$, the wave is gradually slowed down. 
In a 1-D treatment, the phase of the wake during the density transition can be approximated to $\varphi=k_p(z)(z-v_gt)$ in the~leading order. By definition, the local effective oscillation frequency is $\omega_{p,e}=-\partial \varphi/\partial t=v_gk_p(z)$ and the~effective local wave vector is $k_{p,e}=\partial \varphi/\partial z=k_p(z)+(z-v_gt)\partial k_p/\partial z$. Thus, the local phase velocity in the~limit $v_g\rightarrow c$ can be expressed as
\begin{equation}
v_p=\frac{\omega_{p,e}}{k_{p,e}}=v_g\left[1+\frac{1}{k_p}\frac{\partial k_p}{\partial z}(z-v_gt)\right]^{-1}
\approx c\left(1+\frac{\xi}{2n_e}\frac{dn_e}{d\xi}\right)^{-1}
\approx c\left(1-\frac{\xi}{2n_e}\frac{dn_e}{d\xi}\right).
\label{eq:downrampinj}
\end{equation}
A continuous decrease in plasma density results in a reduction of the wave velocity, and this effect is multiplied in subsequent buckets. At some point, the fluid velocity of plasma electrons is reached, and the wave breaks. At sufficiently large distances behind the driver injection of electrons will always occur, as long as the wave is not disturbed by other mechanisms. This injection scheme is often referred to as down-ramp injection \cite{bib:Bulanov1998}.
For example, in a density ramp with $L_{tr}=\frac{n_e}{dn_{e}/dz}=4\lambda_p$, the  electrons with an initial velocity $v_e=c/3$ ($\gamma_e\approx1.1$) will be injected for $v_e=v_p$ at the position $\xi_i=\left(2\frac{c}{v_e}-1\right)\frac{n_e}{dn_e/dz}=16\lambda_p$ behind the driver.
Experimentally this has been demonstrated in the gradient at the rear of a gas jet \cite{bib:Geddes2008, bib:Gonsalves2011} or by using a transverse laser to create a down ramp \cite{bib:Faure2010}.
However, long electron density ramps have two major drawbacks: a) The prolonged injection process will lead to a large energy spread, and 
b) the permanent increase in plasma wavelength causes quick dephasing.

In \textit{sharp density transitions} on a scale length $L_{tr}<\lambda_p$, with a rapid decline from $n_{e,1}$ to $n_{e,2}$, the change of the plasma wavelength $\frac{\Delta \lambda_p}{\lambda_p}\!=\!\frac{\lambda_{p,1}\!-\!\lambda_{p,2}}{\lambda_{p,1}}$ and phase velocity $\frac{\Delta v_p}{v_p}\!=\!\frac{v_{g,1}\!-\!v_{g,2}}{v_{g,1}}$ can be derived as \cite{bib:Buck2011b}
\begin{equation*}
\frac{\Delta \lambda_p}{\lambda_p}\!=\!1\!-\!\sqrt{\frac{n_{e,1}}{n_{e,2}}}\!\approx\!-\frac{\alpha_T-1}{2},\;\;\;\;\;\;\;\; \;\;\;\;\;\;\;\;
\frac{\Delta v_p}{v_p}\!\approx\!1\!-\left(1\!-\!\frac{n_{e,2}}{2n_{e,c}}\right)\left(1\!+\!\frac{n_{e,1}}{2n_{e,c}}\right)\!\approx \!\frac{\alpha_T-1}{2}\frac{n_{e,2}}{n_{e,c}},
\end{equation*}
with the density ratio $\alpha_T= n_{e,1}/n_{e,2}$ and $\alpha_T\gtrsim1$. The priciple is shown in Fig. \ref{fig:shockfront_injection}. As $\lambda_p$ increases almost instantly, the velocity of the plasma wave only increases marginally. Consequently, the velocity of the first wakefield peak suddenly stops, before starting to move again with the new phase velocity. A significant fraction of the electrons, constituting the first wakefield peak before the density transition {\larger\textcircled{\smaller[2]1}} (drawn in gray dashed lines) are now promoted to an accelerating phase after the density transition ($<\SI{10}{\micro\meter}$) {\larger\textcircled{\smaller[2]2}}, as indicated by the yellow area. The injected electrons are all located at a similar phase of the wake with similar initial energy, exposing them to the same accelerating field. Therefore, they will gain similar energy resulting in a quasi mono-energetic beam.

This mechanism has first been proposed numerically \cite{bib:Suk2001,bib:Brantov2008}, and been experimentally verified to produce stable, quasi-monoenergetic electron bunches \cite{bib:Schmid2010, bib:Buck2013}. Here, the gas density profile from a~supersonic gas nozzle was modified by a razor blade creating a narrow shock front with the desired sharp drop in the density profile. Other groups have also reported on the production of monoenergetic beams by using similar approaches, e.g. using wires for the creation of the shock \cite{bib:Burza2013,bib:Brijesh2012}.

The simple implementation of this scheme also favors potential applications, such as a quasi-monochromatic all-optical Thomson x-ray source \cite{bib:Khrennikov2015}. 

\subsubsection{Optical Injection}
\label{ch:th_optical_injection}
\begin{figure}
    \subfigure[\textbf{Beat wave injection} of a pump pulse with$a_{0,1}=1.5$ and an injection pulse with $a_{0,2}=0.8$ (\textbf{top}) and $a_{0,2}=0.2$ (\textbf{bottom}) for circular polarized pulses with a duration of $\SI{35}{fs}$. In the beat wave (bold green line) the fluid/untrapped electrons (dotted black/ dashed blue lines) can gain enough energy ($W_{beat}^{a_{0,2}=0.8}\approx \keV{300}$ and $W_{beat}^{a_{0,2}=0.2}\approx \keV{170} $) to cross the separatrix (purple line) and to be injected on trapped orbits (red line). The yellow area corresponds to the injection volume which determines the injected charge as well as the final energy spread.]
    {\includegraphics[width=0.47\textwidth]{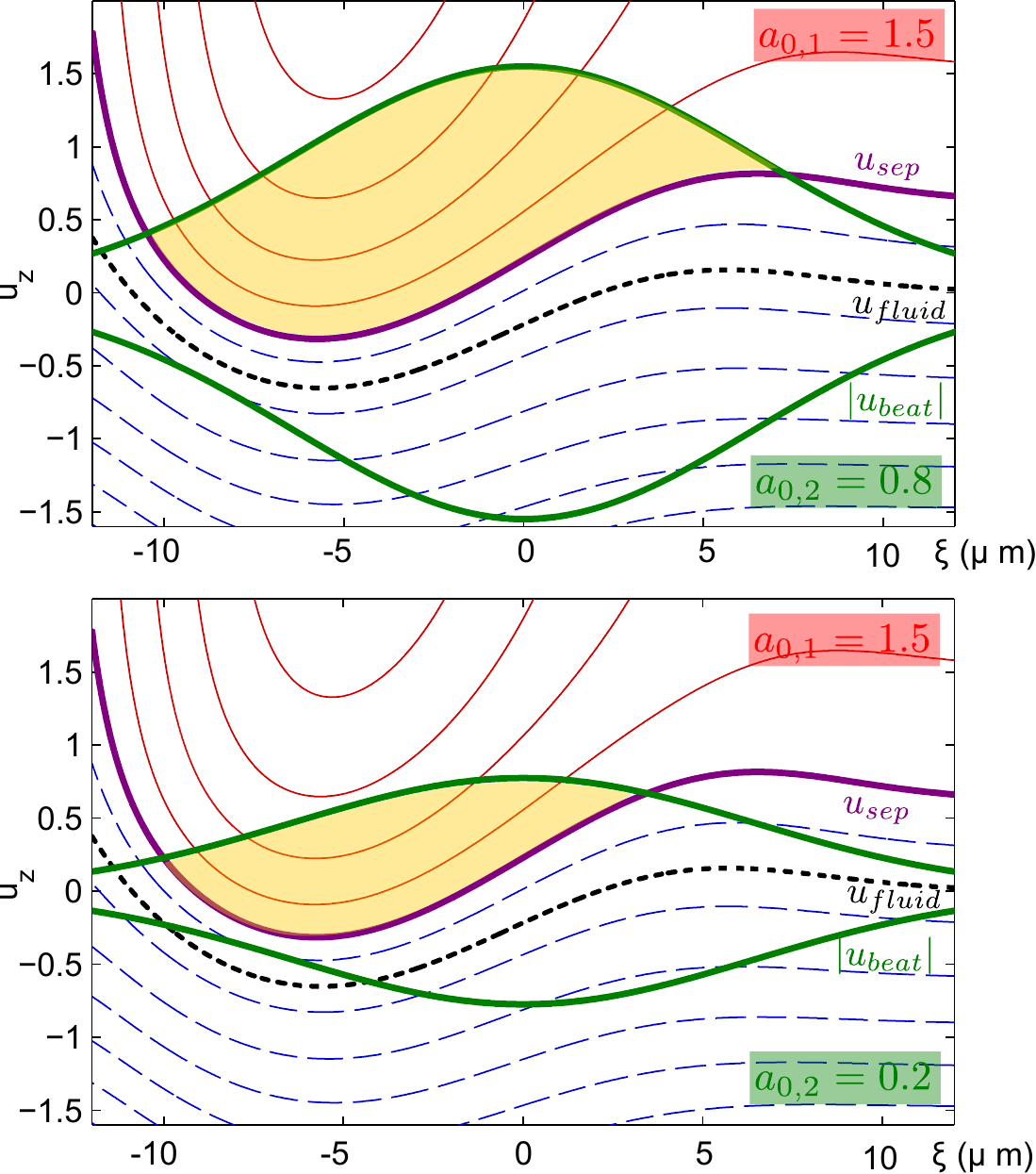}
        \label{fig:separatrix_beatwave}
        }
    \hspace*{0.01\textwidth}
    \subfigure[\textbf{Ionization injection}. For high-$Z$ gases inner shell electrons are ionized at the peak of the laser. (\textbf{center:} the green dashed lines mark the barrier-suppression ionization (BSI) values for nitrogen). The ionized electrons are born with $u_{z}(\xi_{ion})\approx 0$ (dashed black line) and can be trapped by following the green orbits in the phase space. By adjusting the~intensity of the laser pulse the injection region (marked in yellow)the final energy spread $\Delta W$ can be reduced, as shown on the \textbf{top} for $a_0=2.65$ with $\Delta W/W\approx70\%$ and on the~\textbf{bottom} for $a_0=2.25$, $\Delta W/W\approx30\%$.]
    {\includegraphics[width=0.48\textwidth]{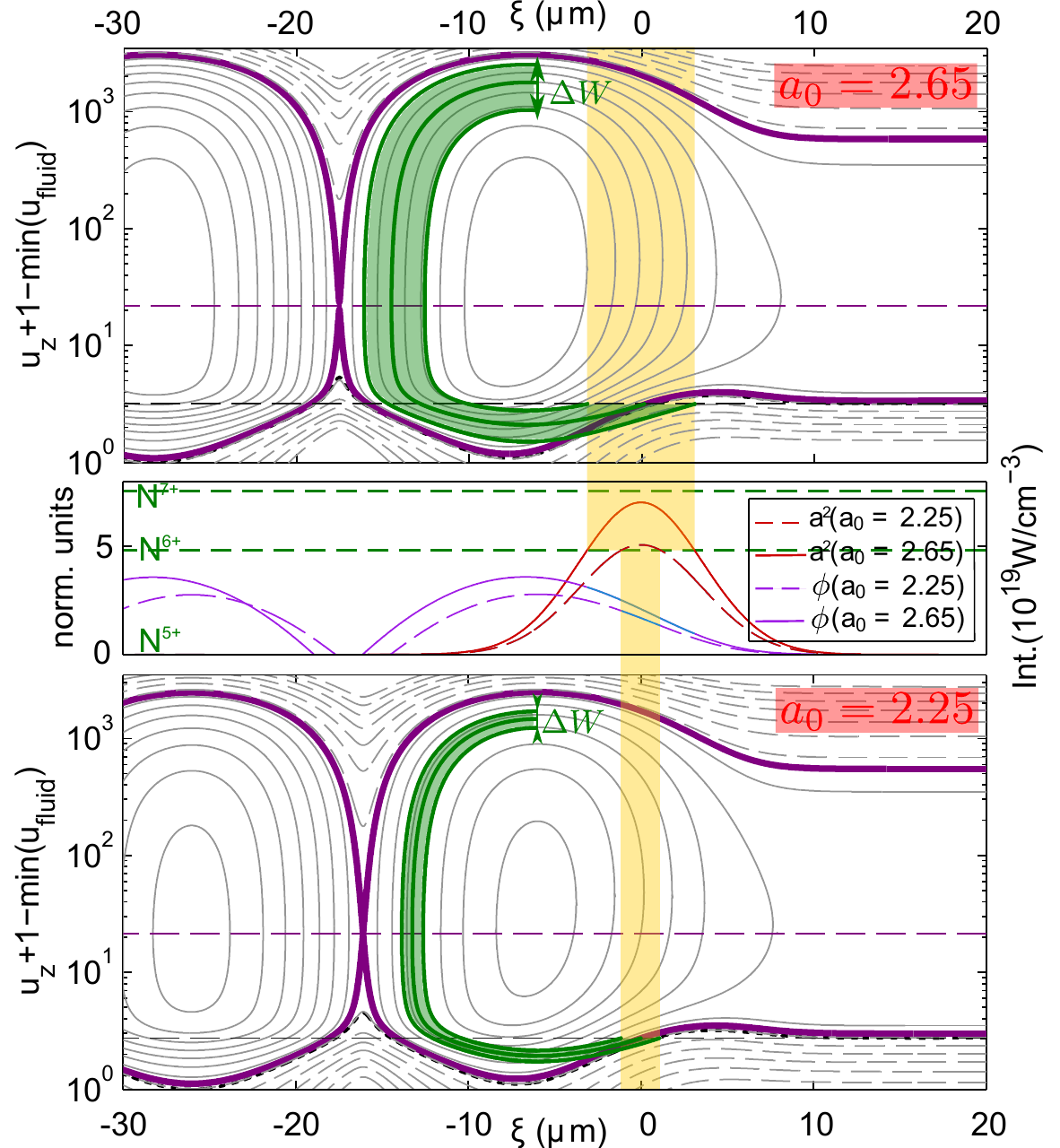}
        \label{fig:ionization_injection}
    }
    \caption
[\textbf{Injection based on density transition and ionization} \newline
    evaluation a:	\textbackslash Th LaserPlasma\textbackslash CP Principle.m \newline
    figure file  a:		\textbackslash Th LaserPlasma\textbackslash CPInj Principle final.pdf/svg          \newline
    evaluation b:	 \textbackslash Th LaserPlasma\textbackslash IonizationInj.m  \newline
    figure file  b:	 \textbackslash Th LaserPlasma\textbackslash Ion Inj Principle final.svg/pdf ]     
    {\textbf{Phase space orbits} demonstrating the principle  for optical injection (\textbf{left}) and ionization injection (\textbf{right}) in 1D  at a plasma density of $n_{e,0}=\SI{5e18}{\per\cubic\cm}$.}
    \label{fig:Phase_space_beatwave_ionization}
\end{figure}

Another possibility to realize a controlled, localized injection relies on two colliding laser pulses. The~first (drive) pulse creates a plasma wave below the self-injection threshold, while the second weaker (injection) pulse counterpropagates and beats with the driver at a position depending on the relative delay. This beat wave causes a localized and instantaneous heating of background electrons, some of which are then injected into the driver's wake \cite{bib:Esarey1997,bib:Fubiani2004}.
Assuming two counterpropagating, circularly polarized laser beams
\begin{align*}
a_{1}\!\!=\!\!\frac{a_{0,1}}{\sqrt{2}}\left(\cos(k_Lz\!\!-\!\!\omega_Lt)\vec{e_x}\!\!+\!\!\sin(k_Lz\!\!-\!\!\omega_Lt)\vec{e_y}\right),\;\;\;\;\;\;\;\;
a_2\!\!=\!\!\frac{a_{0,2}}{\sqrt{2}}\left(\cos(k_Lz\!\!+\!\!\omega_Lt)\vec{e_x}\!\!-\!\!\sin(k_Lz\!\!+\!\!\omega_Lt)\vec{e_y}\right),
\end{align*}
the Hamiltonian (Eq.~(\ref{eq:Hamiltonian})) for the beatwave can be expressed as
\begin{equation*}
H_{beat}=\sqrt{1+u_\perp^2+u_z^2}=\sqrt{1+(a_1+a_2)^2+u_z^2},
\end{equation*}
where $(a_1+a_2)^2=\frac{a_{0,1}^2+a_{0,2}^2}{2}+a_{0,1}a_{0,2}\cos(2k_Lz)$. Note that the Hamiltonian is independent of time. For the initial condition ($u_z=0$ for $z=0$), the beat wave separatrix is
\begin{align*}
u_{beat,sep}&=\pm \sqrt{a_{0,1}a_{0,2}(1-\cos(2k_L z))},  &
u_{beat, max}&=\sqrt{2a_{0,1}a_{0,2}},		&
u_{beat, min}&=-\sqrt{2a_{0,1}a_{0,2}}		
\end{align*}
with the maximum and minimum values $u_{beat, max}$ and $u_{beat, min}$, respectively. Figure \ref{fig:separatrix_beatwave} displays the~trapped and untrapped electron orbits in the phase space generated by a pump pulse with $a_{0,1}=1.5$ and $a_{0,2}=0.8,0.2$, respectively. The ponderomotive force of the beat wave $F_{beat}\propto 2k_La_{0,1}a_{0,2}$ promotes background electrons from the fluid orbits into trapped orbits. The maximum momentum gain is represented by the beat wave separatrix (green solid line). The yellow area marks the accessible phase space in the wakefield for electrons heated by the beat wave \cite{bib:Fubiani2004}
\begin{align*}
u_{beat,max}(\xi)&\gtrsim u_{sep}(\xi), & u_{beat,min}(\xi)&\lesssim u_{fluid}(\xi) .
\end{align*}
The energy gain of the electrons in the beat wave is given by $ W_{beat}=m_ec^2[({1+u_{beat}^2})^{\sfrac{1}{2}}-1]$ $\approx \keV{300}$ for $a_{0,2}=0.8$ and $\approx \keV{170}$ for $a_{0,2}=0.2$, respectively. A lower colliding pulse intensity $a_{0,2}$ reduces the injection phase space resulting in a narrower energy spread, but also reduces the~injected charge \cite{bib:Rechatin2009, bib:Faure2006}. 
Most colliding experiments are performed with linear polarization since this leads to more efficient heating. A full description of optical injection, including the consequences of the heating on the dynamics of the plasma wave, the laser pulse evolution and its influence on the wake formation is beyond the reach of analytical treatment \cite{bib:Davoine2008}, which commonly overestimate, e.g., the injected charge, because of their failure to describe the wakefield inhibition during beating \cite{bib:Rechatin2007}. However, optical injection provides good control over the beam parameters: since the beat wave only exists during the intersection of both laser pulses, the injection is very localized, resulting in quasi-monoenergetic beams. A variable delay between both pulses allows to select the injection position, acceleration distance and final energy, while the amplitude and polarization of the colliding beam controls the the number of trapped electrons and their energy spread (determined by the volume of the phase space and beam-loading) \cite{bib:Faure2006, bib:Malka2009}. 
The disadvantage of optical injection lies in the difficult setup required for splitting, synchronizing and colliding both laser pulses. A direct comparison of optical and shockfront injection can be found in  \cite{bib:Wenz2019}.

\subsubsection{Ionization Injection}
While downramp and optical injection transport electrons across the separatrix boundary for injection  (or vice versa), another way of injection relies on "creating" electrons in already trapped orbits. Since the Barrier-Suppression Ionization (BSI) threshold of inner-shell electrons in medium-$Z$ atoms (e.g., Oxygen, Nitrogen) is comparable to the peak laser field, a small amount of such dopant gas in a Helium or Hydrogen target gas causes ionization events close to the laser peak. An example for $N^{6+}$ is shown in Fig. \ref{fig:ionization_injection}. Created on axis in the wakefield, these electrons have different momentum compared to the~fluid electrons: they are ionized at the position $\xi_{ion}$, at rest ($u_z(\xi_{ion})=0$) (while fluid electrons at this point travel backwards) and close to the laser peak ($a(\xi_{ion})\simeq 0$). The Hamiltonian yields
\begin{equation*}
H_{ion}=1-\phi(\xi_{ion}).
\end{equation*}
If the acceleration field allows the new-born electrons to reach the phase velocity of the wake during before they are overtaken, they are trapped. 
The trapping condition depends on the BSI threshold $a(\xi_{ion})>a_{OTBI}$ for the individual ionization state and by the separatrix, $H_{ion}<H_{sep}$. In Figure~\ref{fig:Phase_space_beatwave_ionization}(b)  the~orbits of the trapped electrons are marked green. The trapping region (marked in yellow) can be adjusted by the intensity of the laser pulse. It defines the final energy spread $\Delta W$, as shown for $a_0=2.65$ on the top and $a_0=2.25$ on the bottom of the figure. 
Under the further assumption that the electrons are released only at the maximum of the laser pulse with zero momentum, $a(\xi_{ion})\approx0$, 
the trapping condition ($H_{ion}<H_{sep}$) can be approximated to the inequality \cite{bib:Chen2012} 
\begin{equation*}
1-1/\gamma_p\leq\phi(\xi_{ion})-\phi_{min}\leq\phi_{max}-\phi_{min}.
\end{equation*}
This sets a minimum for the required laser intensity 
\begin{equation*}
1-\frac{1}{\gamma_p} \leq a_0^2+\frac{a_0^2}{a_0^2+1}.
\end{equation*}
For a squared pulse (Eqs. (\ref{eq:max_electric_field}) and (\ref{eq:separatrix_phi})) the inequality is satisfied at typical gas densities ($\gamma_p>10$) for $a_0>0.7$. As the main consequence, ionization induced trapping requires relativistic laser intensities. Experimental evidence of this scheme is given by \cite{bib:Pak2010}, where an efficient injection of electrons in He:Ar gas mixtures at a vector potential 
 well below the self-trapping regime has been observed.
Although its simple implementation and high-charge output, ionization injection yields broad spectra, because: a) Injection is continuous, as long as the trapping condition is fulfilled, and b) Tunnel ionization smears out the ionization threshold $a_{OTBI}$. For a linearly polarized laser, the tunnel ionization rate can be approximated \cite{bib:Keldysh1965,bib:Esarey1997a} as
\begin{equation*}
\Gamma(\vert E_L\vert)=4\left(\frac{3}{\pi}\right)^{1/2}\Omega_0\left(\frac{W_{ion}}{W_{ion,H}}\right)^{7/4}\left(\frac{E_H}{\vert E_L\vert}\right)^{1/2}
\exp\left(-\frac{2}{3}\left(\frac{W_{ion}}{W_{ion,H}}\right)^{3/2}\frac{E_H}{\vert E_L \vert}\right),
\end{equation*}
where $E_L$ is the laser field and $E_H=\SI{5.2}{\giga\volt\per\centi\meter}$ the ionization field of hydrogen,  $\Omega_0=\alpha_fc/r_B=\SI{4e16}{\per\second}$ is the characteristic atomic frequency, $W_{ion}$ and $W_{ion,H}=\SI{13.6}{eV}$ the ionization energy of the specific ion and of hydrogen, respectively. The fraction of species which is
ionized during a time $\Delta t$ is given by $\Gamma \Delta t$. For example, the ionization probability for $N^{5+}\rightarrow N^{6+}$ in a the presence of a Gaussian pulse $a=a_0\exp(-t^2/\tau_0^2)$ with $t_{FWHM}=\sqrt{2\ln(2)}\tau_0=\SI{28}{fs}$ is $\sim1\%$ for $a_0=1.5$ and increases to $90\%$ for $a_0\sim2.0$, i.e., the threshold for ionization lies in the range between $a_0=1.5-2.0$. 
Accordingly, small fluctuations in laser self-compression and self-focusing  have a direct impact on the injected charge and decrease the overall stability.
In principle, by reducing the laser pulse energy, the energy spread of an electron bunch can be reduced, but so will its charge and energy.
The best experimental results using a $1\%$ additive of nitrogen have demonstrated an energy spread of $20\%$ \cite{bib:McGuffey2010}. Localizing the dopant concentration can further reduce the energy spread, as demonstrated in  \cite{bib:Pollock2011}.

\subsection{The Bubble Regime}
\label{ch:bubble_regime}
The analytical 1D studies presented so far offers a basic understanding of the LWFA process. The lack of analytical 3D models necessitates numerical simulations to capture, e.g., the transverse dynamics or the laser pulse evolution. Commonly, the particle-in-cell method (PIC) is used to study a more complete picture of the laser, wakefield and electron bunch evolution.
 
\begin{figure}
    {\includegraphics[width=0.9\textwidth]
   {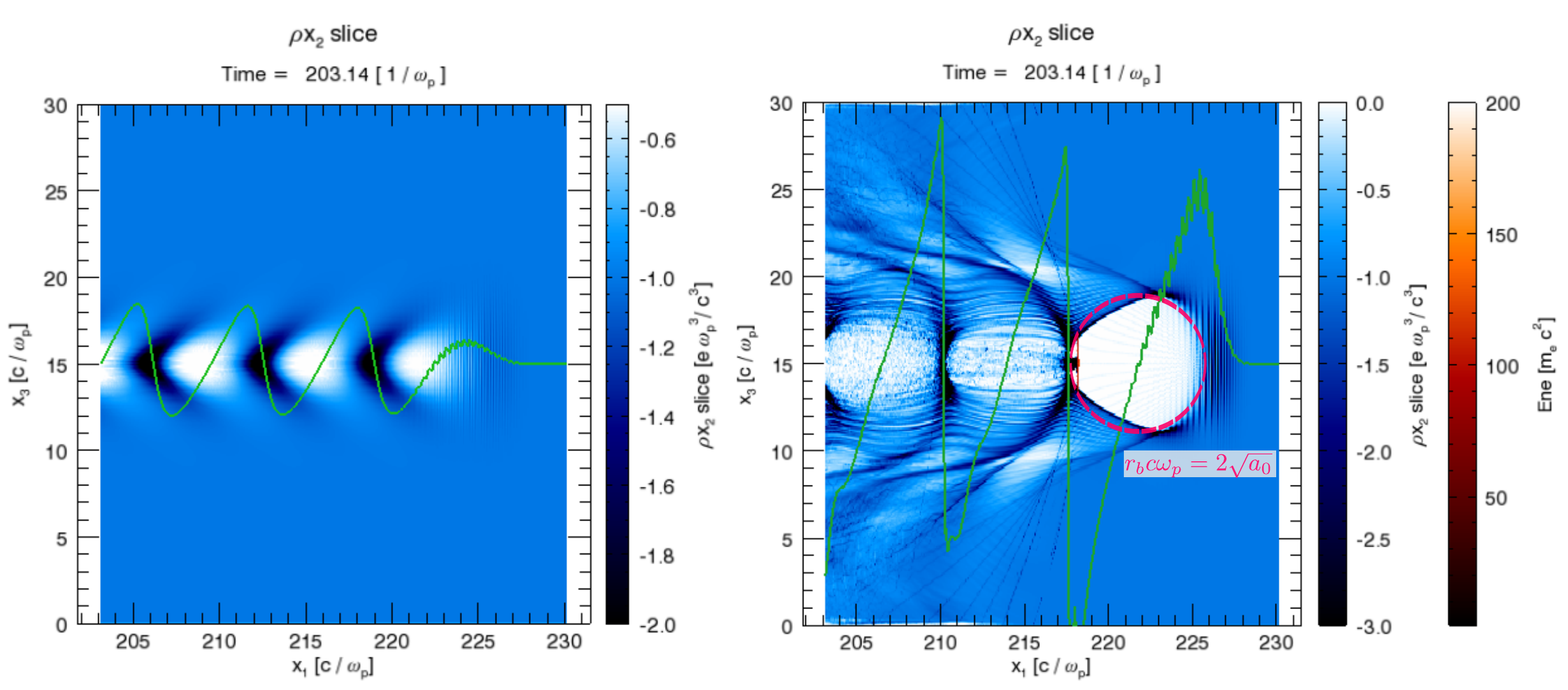}
    \caption
        [\textbf{\textit{Osiris} simulation of LWFA} \newline
    input file a:    \textbackslash Th LaserPlasma\textbackslash Osiris3d a10 \textbackslash os-stdin\newline
    input file b:   \textbackslash Th LaserPlasma\textbackslash Osiris3d a41 \textbackslash os-stdin\newline
    figure file  :		\textbackslash Th LaserPlasma\textbackslash Osiris3d a4and a1.pdf/svg 
]   
{\textbf{\textit{3D OSIRIS} PIC simulation of LWFA acceleration}  The laser ($\lambda_L=\SI{800}{nm},t_{FWHM}=\SI{28}{fs}$), propagating towards the right,  is focused to a spot size of $d_{FWHM}=\SI{9.62}{\micro\meter}$ with an energy of $W_L=\SI{100}{mJ}$ on the left and $W_L=\SI{1.55}{J}$ on the right. The electron density ($n_{e,0}=\SI{5e18}{\per\cubic\cm}$) is shown in blue in the $(y\!\!=\!\!0)$-plane.\\
\textbf{Left:} In the modest relativistic regime ($a_0\approx 1$), the laser pulse (at $x_1 \approx 226$  $c/\omega_p$) excites a weakly nonlinear wake (electric field on axis shown in green). The bubble is not completely void of electrons and the wake's wavefronts are horse shoe-shaped. No self-injection occurs during this simulation before the laser is depleted.\\
\textbf{Right:} In the highly nonlinear regime ($a_0 \approx 4.1$), the simulation parameters correspond to the matched condition $d_{FWHM}\approx2\sqrt{a_0}/k_p=\SI{9.6}{\micro\meter}$. The bubble with radius $r_bc\omega_p=2\sqrt{a_0}$ (magenta circle) is fully developed, the~electron density drops to zero and the created fields exceed the wavebreaking limit. Near the rear of the bubble the onset of transverse self-injection is visible. The energy of the injected electrons is color-coded in black-and-red.}
\label{fig:bubble}
}
\end{figure}

Based on such numerical 3D PIC studies, \cite{bib:Pukhov2002} have found that under optimized conditions and a~strong driver, a very efficient acceleration is possible. Their scheme relies on the relativistic ponderomotive force to radially expel all plasma electrons from the laser path. This forms an electron cavity behind the~laser pulse, where only the ions stay behind. After a distance of $\lambda_{p,rel}$, they pull back the~electrons, whose trajectories cross on axis, forming a density spike. The shape of the electron cavity is approximately spherical, hence the name  ``bubble'' regime, see Fig. \ref{fig:bubble}. In the bubble, the electron density is zero, and this regime is also referred to as ``complete blow-out''.
Experimentally, this regime has been studied extensively since the first observation by \cite{bib:Faure2004}.

\cite{bib:Lu2006},\cite{bib:Kostyukov2004} 
 have developed phenomenological laws for the bubble regime via extensive studies of PIC codes underpinned by simplified analytical studies.
They found that a spherical structure with a~radius given by $r_b\approx2\sqrt{a_0}/k_p$ is formed. The normalized potential and the electric fields inside the~cavity moving at relativistic speed are \footnote{Note that in these 3D wakes the wakefields are electromagnetic in character, not only electrostatic.} 
\begin{align*}
\phi=&\frac{k_p^2}{4}\left(r^2-r_b^2\right) ,   & 
\frac{E_z}{E_{p,0}}=&\frac{k_p}{2}\xi ,   &
\frac{E_{\perp}}{E_{p,0}}=&\frac{k_p}{4}r_\perp,
\end{align*}
where $r^2=\xi^2+r_\perp^2=0$ is the radius from cavity center, $r_\perp^2=x^2+y^2$ is the transverse and $\xi$ the longitudinal position. The radial fields $E_\perp$ are linear in $r_\perp$, and therefore emittance-preserving. The~longitudinal fields are almost linear with a gradient $\vert\partial E/\partial \xi\vert= E_{z,max}/r_b$, with
\begin{align}
E_{z,max}=&\frac{r_bk_p}{2} E_{p,0}, &  E_{z,max}[GV/m]&\approx 96\sqrt{n_{e,0}[10^{18}cm^{-3}]}\sqrt{a_0}.
\label{eq:el_field_in_bubble}
\end{align}
The radial and longitudinal forces acting on the trapped electrons are given by
\begin{align}
\vec{F_z}&=-\frac{1}{2}m_e\omega_p^2\xi \vec{e_z},  &  \vec{F_r}&=-\frac{1}{2}m_e\omega_p^2r_\perp \vec{e_r} .
\label{eq:force_in_bubble}
\end{align}

In the leading half of the bubble ($0<\xi<r_b/2$), the electron bunch is decelerating, while in the back of the bubble it accelerates, with a gradient independent of its radial position. The transverse focusing region for electrons extends over the whole bubble, defocusing only occurs around the on-axis density peak. The focusing fields are of similar magnitude as the longitudinal fields (\SI{100}{\giga\electronvolt\per\meter}).

Injection into the bubble occurs via transverse wave-breaking (cf. Section \ref{ch:self_injection}) at the density spike, at the location of the highest accelerating fields. Due to the large initial deflection by the ponderometive force, electrons forming the spike exhibit a large transverse momentum, resulting in a transverse emittance. The optimum condition for bubble formation and injection can be found by matching the laser intensity, spot size $w_b$, pulse length ($c\tau_0\leq w_b$) and plasma wavelength as shown in Fig. \ref{fig:bubble} \cite{bib:Lu2007}
\begin{equation}
w_b\simeq r_b= 2\sqrt{a_0}/k_p.
\label{eq:bubble_condition}
\end{equation}  
For $a_0>4$ and a focal spot size of $w_b=d_{FWHM}=2\sqrt{a_0}/k_p$, the ion cavity takes the shape of a~perfect sphere. In terms of the critical power $P_{crit}$ (Eq. (\ref{eq:crit_power})) this condition yields $P\simeq (a_0/2)^3P_{crit}$.   
Exemplary, at $n_{e,0}=\SI{5e18}{\per\cubic\cm}$, a peak power of $\SI{42}{TW}$, focused to $a_0>4$ is required.

Between $2<a_0<4$, the blow-out cavity deviates slightly from a spherical shape. 
However, self-focusing and self-compression will match even initially unmatched pulses \cite{bib:Lu2007} to drive a bubble.

Self-guiding losses are minimized when the initial spot size matches the blow-out radius, as experimentally confirmed by \cite{bib:Ralph2009}.
In \cite{bib:Thomas2007} it was found that relativistic self-focusing has a stabilizing effect, allowing the wakefield to evolve to the correct shape even from fluctuating initial parameters. The best performance is achieved for long focal-lengths ($w_b\gtrsim \lambda_p$), and a Rayleigh length larger than the density gradient at the plasma entrace. Then the majority of the pulse energy coalesces into a single optical filament and can be self-guided over distances comparable to the dephasing length. Smaller spots usually result in beam breakup and poor reproducibility. Thus, increasing in the pulse intensity through self-evolution is more effective than a tight focusing geometry. The choice of the spot size (small enough to avoid depletion, large enough to avoid filamentation) is crucial for the production of high-quality beams. 

\subsection{Limitations on Energy Gain}
\label{ch:limitsofLWFA}
The main limitations for LWFA are the "detrimental Ds":  Diffraction, Dephasing, and Depletion.

\subsubsection{Laser Diffraction}
In any focused beam, diffraction will reduce the laser intensity after a certain distance, but self-focusing may balance this over many Rayleigh lengths. The general refractive index (cf. Eq. (\ref{eq:refractive_index_long})) is given by

\begin{equation}
\eta=\frac{ck}{\omega_L}=
\sqrt{1-\frac{\omega_p^2}{\gamma_\perp \omega_0^2}}\approx1-\frac{n_e}{2\left(1+a_0^2\right)n_c} \\ 
\label{eq:refractiveindex}
\end{equation}
For the 3D bubble regime, the self-guiding condition can be reformulated to \cite{bib:Lu2007}
\begin{equation}
a_0=\left(\frac{n_c}{n_e}\right)^{1/5}.
\label{eq:self-guiding}
\end{equation}
If this condition is fulfilled, the laser propagates with a spot size close to  $w_0=2\sqrt{a_0}/k_pr$. In a realistic scenario, however, $a_0$ will change by self-compression and pump depletion, while an initially unmatched laser will also oscillate in size. Despite the complex dynamics that is best treated by numerical simulations, one thing is certain: Once the pulse depletes, it will quickly diffract.

\subsubsection{Electron Dephasing}

Trapped electrons can reach velocities closer to $c$ than the laser group velocity, so they catch up with the laser and "dephase", i.e. enter the decelerating phase of the wake (cf. point {\larger\textcircled{\smaller[2]C}} in Fig. \ref{fig:separatrix}). Consequently, the maximum energy which can be extracted from the wake depends on the (relativistic) plasma wavelength. The distance which an electron can travel before it crosses the zero field, $E_z=0$, in the~lab frame is called \textit{dephasing length} $L_d$. In the linear regime, it corresponds to the distance in which a~relativistic electron ($v_e\approx c$) phase slips by $\lambda_p/4$ with respect to the laser field ($v_g=\eta c$). Only in this phase space region the fields are accelerating and simultaneously focusing, cf. green area in Fig. \ref{fig:linearsolution}. With the~refractive index $\eta$ (Eq. (\ref{eq:refractiveindex})) the dephasing length can be derived to
\begin{align*}
\left(c-v_g\right)\frac{L_{d}}{c}=&\frac{\lambda_p}{4} & \Rightarrow & &
L_{d}=&\frac{\lambda_p}{2}\frac{\omega_L^2}{\omega_p^2}
\end{align*}
In the nonlinear regime $a_0\gg 1$, the increase in the plasma wavelength has to be included. For a linearly polarized, square pulse (Eq.(\ref{eq:nonlinear_plasmawavelength})) the dephasing length at an arbitrary intensity is given accordingly by 
\begin{align}
L_{d}\approx
\gamma_p^2 \lambda_{p} \times \begin{cases} \frac{1}{2}  &\text{for } a_0^2 \ll 1\\ \frac{1}{\pi} a_0 & \text{for } a_0^2 \gg 1\\ \end{cases}
\label{eq:dephasing1D}
\end{align}
$L_{d}$ scales inversely with the electron density $\propto n_e^{\sfrac{-3}{2}}$, 
and can therefore be mitigated by operation at lower densities, albeit at longer acceleration length. For typical densities of $n_{e,0}\approx\SI{5e18}{\per\cubic\centi\meter}$ and $a_0=2$, the dephasing length is usually $L_d\sim\SIrange[range-phrase=-,range-units = single]{3}{4}{mm}$.
Using a phenomenological approach, \cite{bib:Lu2007} derived for a \SI{3}{D} nonlinear regime a dephasing length of (cf. Table \ref{tab:scalinglaws})
\begin{equation}
L_{d}^{3D}=\frac{4}{3}\frac{\sqrt{a_0}}{k_p}\frac{\omega_L^2}{\omega_p^2}.
\label{eq:3d_dephasing}
\end{equation}
In positive density gradients, dephasing can be mitigated by keeping the electrons at a fixed phase, if the~phase slippage of the electrons is compensated by a matched reduction of $\lambda_p $\cite{bib:Katsouleas1986,bib:Rittershofer2010}.

\subsubsection{Pump Depletion}
\label{ch:pump_depletion}
\begin{figure}
    \subfigure[\textbf{Optimized beam charge} color coded on a logarithmic scale as given by Eq. (\ref{eq:opt_charge}). The charge is roughly constant for the matched beam parameters around $Q_{opt}\approx \SI{350}{pC}$. As example, for $a_0=4.3$ the dephasing length (blue dotted dashed line) and the depletion length (blue dotted line) for the 3D nonlinear regime are displayed (right axis).]
    {\includegraphics[width=0.47\textwidth]{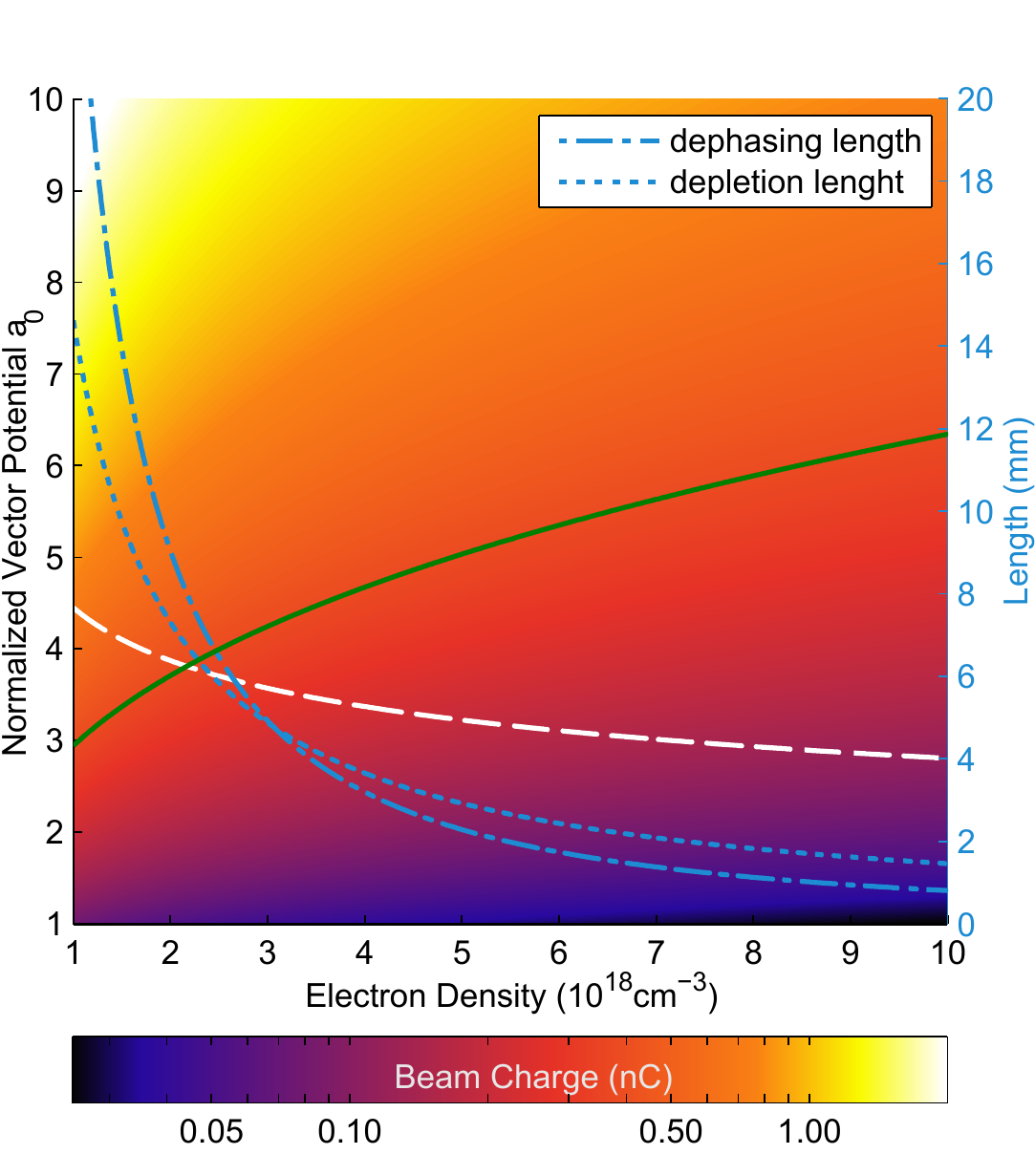}
        \label{fig:optimumCharge}
    }
        \hspace*{0.02\textwidth}
        \subfigure[\textbf{Maximum electron energy gain.} The highest electron energies are given for the lowest electron densities, as long as the self-guiding condition is satisfied, e.g.: $W_{Lu}=\MeV{510}$ for $n_{e,0}=\SI{5e18}{\per\cubic\cm}$ and $a_0=4.3$. The~corresponding maximum spot size for the ATLAS parameters $d_{FWHM}=f(a_0,W_L,t_{FWHM})$ is given by the dashed blue line and top axis.]
    {\includegraphics[width=0.47\textwidth]{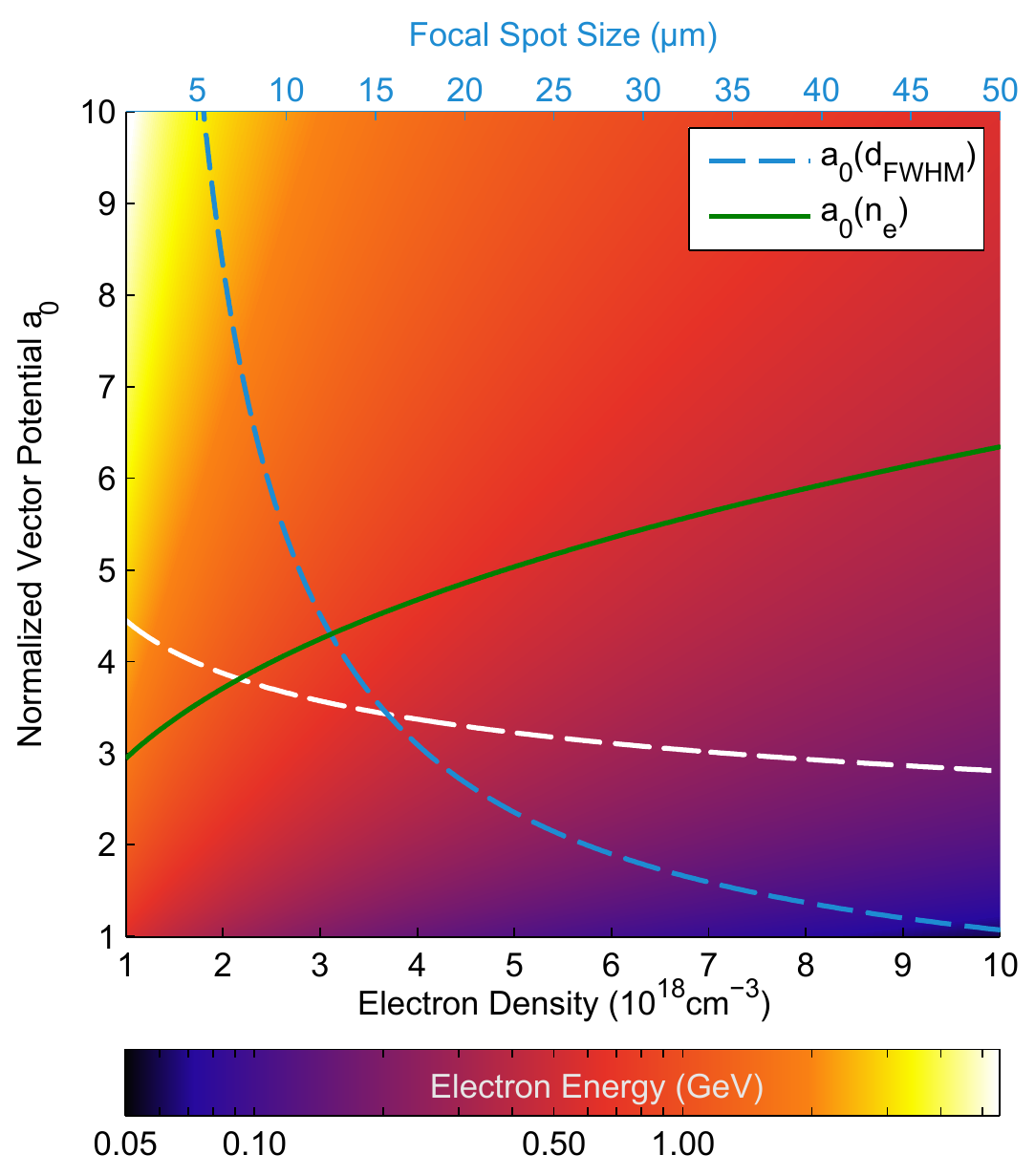}
        \label{fig:maximumGain}
        }
    \caption
        [\textbf{Electron output parameters in the matched condition} \newline
    evaluation   :	 \textbackslash Th LaserPlasma\textbackslash MaximumGainandLimits.m \newline
    figure file a:	 \textbackslash Th LaserPlasma\textbackslash Optimum Charge final.pdf/svg \newline
    figure file b:		\textbackslash Th LaserPlasma\textbackslash Maximum Gain final.pdf/svg  ]    
    {\textbf{Electron output parameters in the matched condition} $k_pr_b=2\sqrt{a_0}$. The white dashed line represents the laser self-guiding condition (Eq. (\ref{eq:self-guiding})), i.e., the area below the white line can be only accessed via external guiding. The solid green line shows the matched condition ($w_0k_p=2\sqrt{a_0}$) for a laser with $W_L\!\!=\!\!\SI{1.5}{J}$, $t_{FWHM}\!\!=\!\!\SI{28}{fs}$ assuming a Gaussian pulse with $w_b=d_{FWHM}$.}
    \label{fig:Bubble_charge_and_energy}
\end{figure}

During its propagation, the laser's energy is transferred to the plasma wake until it is depleted and acceleration ends. 
The characteristic length for pump depletion $L_{pd}$ can be estimated by comparing the~energy density in the wake, $u_W=\frac{1}{2}\epsilon_0 E_z^2$, contained in the volume $V=\pi w_0 L_{pd}$, and the laser energy density, $u_L=\frac{1}{2}\epsilon_0 E_0^2$, contained in the volume $V=\pi w_L c \tau_L$. With $a_0=eE_0/cm_e \omega_L$ (Eq.~(\ref{eq:normalized_vector_potential})) and the relation for linear wakes $E_z/E_{p,0}\approx a_0^2$ (Eq.~(\ref{eq:linear_Ez})), the pump depletion is
\begin{equation*}
L_{pd}=\frac{\omega_L^2}{\omega_p^2} \frac{c \tau_L}{a_0^2}\approx\frac{\lambda_p^3}{\lambda_L^2 a_0^2}.
\end{equation*}
In the relativistic case ($a_0\gg1$) for 1D square pulses with $c\tau_L\approx \lambda_{p,rel}/2$, the scaling of $E_z$ (\ref{eq:max_electric_field}) yields:
\begin{align}
L_{pd}\approx
 \gamma_p^2 \lambda_{p} \times \begin{cases} \frac{1}{a_0^2}  &\text{for } a_0^2 \ll 1\\ \frac{a_0}{\pi} & \text{for } a_0^2 \gg 1\\ \end{cases}
\label{eq:pumpdepletion1D}
\end{align}
In the 3D nonlinear regime, \cite{bib:Decker1996} estimates the pump depletion from the laser's head erosion as it drives the wake. This picture is confirmed by simulations in \cite{bib:Lu2007}, which gives the etching velocity as $v_{etch}\simeq c\omega_p^2/\omega_L^2$. This means that the pulse is fully depleted after a length
\begin{equation}
L_{pd}^{3D}=\frac{c}{v_{etch}}ct_{FWHM}\simeq\frac{\omega_L^2}{\omega_p^2}ct_{FWHM}.
\label{eq:3dpump_depletion}
\end{equation} 
Figure \ref{fig:optimumCharge} displays the \SI{3}{D} dephasing and depletion lengths as blue dashed- dotted and blue dashed lines, respectively. Although both would be equal at $n_e = 3 \times 10^{18}$ W/cm$^3$ and $a_0 \approx 3.3$, indicating optimum energy transfer, the plot indicates that the laser would not be self-focused and also not matched to the plasma (white and green lines, respectively). A shifted operation point (e.g., $n_e = 2.5 \times 10^{18}$~W/cm$^3$ and $a_0 \approx 4.1$) would fulfil these conditions, albeit with depletion now happening before dephasing. Therefore, this plot can help in selecting the best working point for a given set of laser parameters. 
\\

\subsubsection{Beam Loading}
\label{ch:beamloading}
So far, only single test electrons in the plasma fields have been considered. A highly charged bunch, however, will modify ("beam-load") the electric field of the plasma wave, and therefore the acceleration process. Typically, a beam loaded wakefield will manifest itself in a decrease in energy gain and a~modified energy spread. In a nonlinear 1D approach, Poisson's equation has to be amended by a term describing the bunch density distribution $n_b\left(\xi\right)$
\begin{equation*}
\frac{\partial^2 \phi}{\partial \xi^2}=k_p^2\left(\frac{n_e}{n_{e,0}}-1+\frac{n_b}{n_{e,0}}\right),
\end{equation*}
and the solution for the PDE, analogous to the Eq. (\ref{eq:poissonincomov}) is given by \cite{bib:Schroeder2006}
\begin{equation}
\frac{\partial^2 \phi}{\partial \xi^2}=k_p^2 \gamma_p^2 \left[\beta_p\left(1-\frac{\gamma^2}{\gamma_p^2 \left(1+\phi\right)^2}\right)^{-\sfrac{1}{2}}-1 \right]+k_p^2\frac{n_b\left(\xi\right)}{n_{e,0}}.
\label{eq:PDEsolution_beamloading}
\end{equation}
\begin{figure}
\begin{center}
\includegraphics{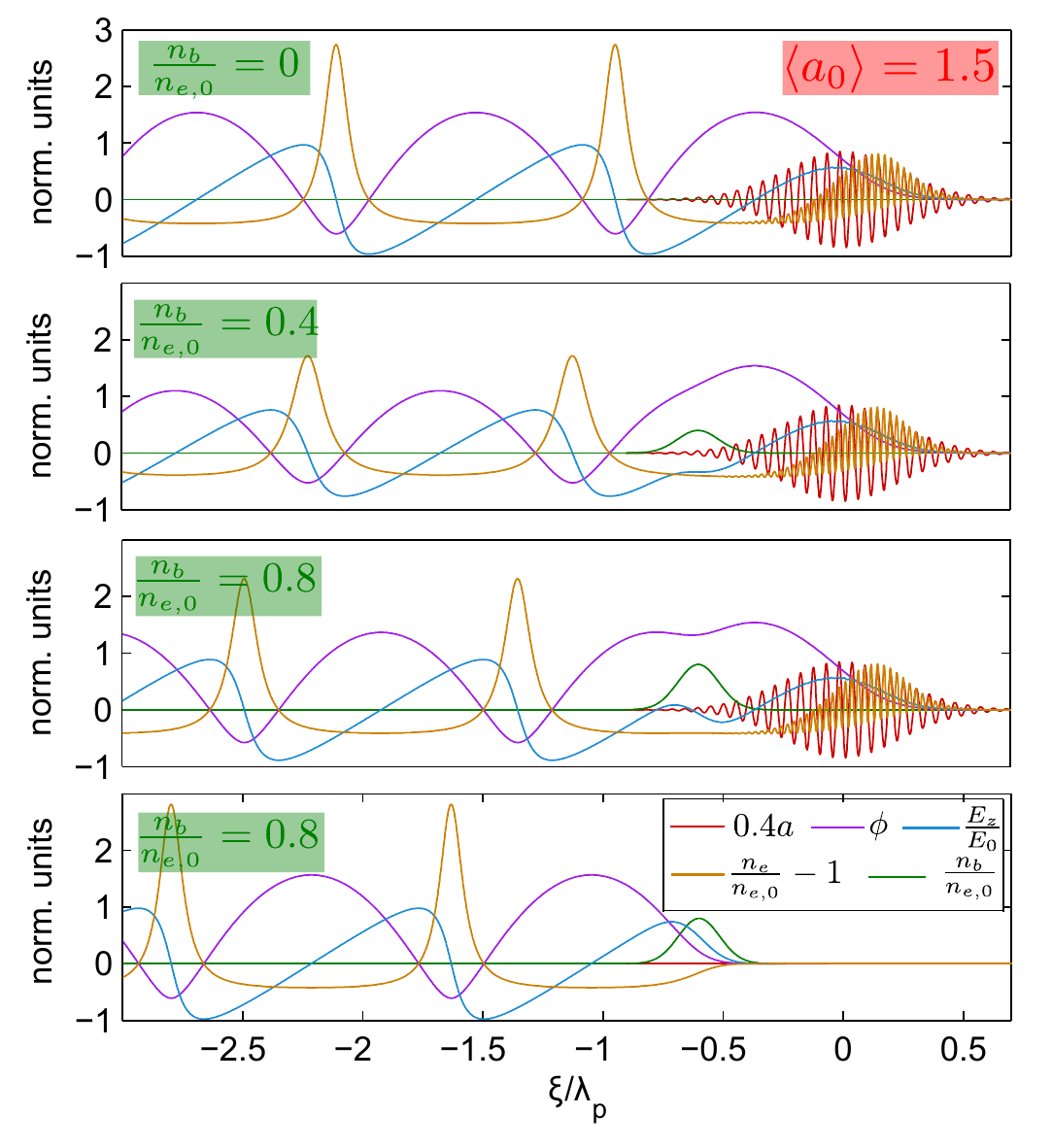}
\caption{ \textbf{Effect of beam loading.} Normalized 1D wakefield quantities , calculated with Eq.~(\ref{eq:PDEsolution_beamloading}). \textbf{top}: laser driver (linearly polarized, peak $a_0=
    2.1$, $t_{FWHM}=\SI{20}{fs}$), \textbf{bottom}: electron driver ($n_b/n_{e,0}=0.8$, $\tau_b=\SI{10}{fs}$). \textbf{two central plots}: mixed driver, laser at $\xi = 0$, electron bunch at $\xi=-0.6\lambda_p$. Peak bunch density $n_b/n_{e,0}=0.4$ and $n_b/n_{e,0}=0.8$, for the second and third panel, respectively. The bunch charge modifies both the potential and the~longitudinal electric field and lengthens the cavity. Electron density for all cases: $n_{e,0}=\SI{5e18}{\per\cubic\cm}$. }
\label{fig:beamloading}
\end{center}
\end{figure}

The bunch's self-fields will superimpose and modify the accelerating field. Figure  \ref{fig:beamloading} shows the beam loading effect upon the wakefield. A Gaussian electron bunch ($t_{b,FWHM}\!=\!\SI{10}{fs}$), is placed at the~$\xi\!=\!-0.6 \lambda_p$ behind the drive laser $\langle a_0 \rangle=1.5$. In the top and bottom panel, a purely laser (bunch) driven wake is shown, respectively, while the second and third panel display two cases with $n_b/n_{e,0}\!=\!0.4$ and $n_b/n_{e,0}\!=\!0.8$. In both cases, the electric field and wave potential are perturbed by the charge to a~varying degree, manifesting itself in a modified peak field and an elongated first wakefield bucket.

However, this superposition can also have a positive side-effect.  For the case $n_b/n_{e,0}\!=\!0.4$ shown in the second panel, the accelerating field (blue line), normally decreasing towards smaller values of $\xi$, stays constant over the duration of the electron bunch. This means that all electron in the bunch are accelerated by the same field, irrespective of their phase in the wake. This reduces chirp and energy spread of the accelerated bunch \cite{bib:Tsung2004}.
In addition, via the reduction of the wakefield amplitude, beam loading can terminate self-injection once a high-charge bunch occupies the first bucket
 \cite{bib:Rechatin2009a, bib:Couperus2017}.\\
The maximum injected charge can be estimated by assuming a sphere with a radius $r_b=2\sqrt{a_0}/k_p$, carrying the total ionic charge $Q_{tot}=\frac{4}{3}r_b^3n_{e,0}e\propto a_0^{3/2}n_{e,0}^{-1/2}
$.
The optimum injected charge $Q_{opt}$ has been derived 
for a matched laser driver by
\cite{bib:Tzoufras2008} to
\begin{align}
Q_{opt}=&\frac{\pi\epsilon_0m_e^2c^4}{e^2}\left(\frac{k_pr_b}{2}\right)^4\frac{1}{E_z}  &\xRightarrow[]{E_z=r_bk_pE_{p,0}/2} & 
&Q_{opt}=&\frac{\pi c^3}{e^2}\left(m_e^3\epsilon_0^3{a_0^{3}}/{n_{e,0}}\right)^{1/2},
\label{eq:opt_charge}
\end{align}
which results in $Q_{opt}\approx Q_{tot}/10$.
For typical parameters, $a_0\sim 2-4$, $n_{e}=\SI{5e18}{\per\cubic\cm}$, the injected charge is $Q_{opt}\sim\SI{350}{\pico\coulomb}$.
In Figure \ref{fig:optimumCharge}, $Q_{opt}$ is plotted with respect to $a_0$ and $n_{e,0}$. The charge can be increased for higher laser pulse intensities and larger bubble structures, i.e., lower electron densities.
The white dashed line represents the self-guiding condition (Eq. (\ref{eq:self-guiding})), which has to be fulfilled in order to sustain the laser intensity over the desired length.

\vspace{-0.01\textwidth}
\subsection{Scaling Laws}
\label{ch:scaling_laws}
The total energy gain $\Delta W$ of an electron determined by the wake's longitudinal electric field $E_z(z)$ and the distance $L_{acc}$
\begin{equation*}
\Delta W=-e \int_0^{L_{acc}} E_z\left(z\right)\, \mathrm{d}z.
\end{equation*}
Neglecting diffraction, dephasing in the linear regime and pump depletion in the nonlinear regime limit $L_{acc}$. In 1D, $\Delta W$  is given by $L_d$ or $L_{pd}$ (Eqs.~(\ref{eq:dephasing1D}) and (\ref{eq:pumpdepletion1D})) and the maximum field (Eqs. (\ref{eq:linear_Ez}) and (\ref{eq:max_electric_field})):
\begin{align*}
\Delta W\approx e\frac{E_z}{2}L_{acc}=\begin{cases} e(E_0a_0^2)L_d= m_ec^2a_0^2 \gamma_p^2\pi/2  &\text{for } a_0 \ll 1\\  
e(E_0a_0/2)L_{pd}= m_ec^2a_0^2 \gamma_p^2 &\text{for } a_0 \gg 1\\ \end{cases}.
\end{align*}
Both regimes scale equally with $a_0$ and $n_e$, and dephasing can be overcome by longitudinal density tapering.
However, operation the nonlinear regime has the advantage of higher accelerating gradients, therefore, shortening the acceleration length and often avoiding the need for guiding.
In the nonlinear 3D regime, the energy gain for a strong driver is limited by dephasing and can be approximated via Eq.~(\ref{eq:el_field_in_bubble})
\begin{equation}
W_{el}^{Lu}=e\langle E_z\rangle L_d^{3D}=\frac{2}{3} a_0 m_ec^2 \frac{n_{c}}{n_{e,0}}.
\label{eq:max_energy_gain_3D}
\end{equation}
 Table \ref{tab:scalinglaws} summarizes some LWFA-scaling rules from the literature \cite{bib:Pukhov2002,bib:Lu2007}.
\begin{table}
\small
\centering
\caption{Scaling rules for LWFA in the linear and nonlinear 1D and 3D  regime as given by \cite{bib:Esarey2009,bib:Lu2007,bib:Pukhov2004}}
\begin{tabular}{|l|l|l|l|l|l|l|l|}
\hline \hline
& \pmb{$a_0$} & \pmb{$w_0$}  & \pmb{$L_d$} & \pmb{$L_{pd}$} & \pmb{$\lambda_p$} & \pmb{$E_z/E_{p,0}$} & \pmb{$\Delta W/m_ec^2$}\\
\hline \hline

Linear & $<1$  &  $\frac{2\pi}{k_p}$  		&  $\frac{\pi}{k_p}\frac{\omega_L^2}{\omega_p^2}$   &  $\frac{c\tau_L}{a_0^2}\frac{\omega_L^2}{\omega_p^2}$ 			&  $\frac{2\pi}{k_p} $ & $a_0^2$ & ${\pi} a_0^2\frac{\omega_L^2}{\omega_p^2}$ \\
\hline
1D NL & $>1$  &  $\frac{2\pi}{k_p}$  		&  $\frac{2a_0}{k_p}\frac{\omega_L^2}{\omega_p^2}$   &  $a_0\frac{2}{k_p}\frac{\omega_L^2}{\omega_p^2}$ 			&  $\frac{4a_0}{k_p} $ & $a_0/2$ & $ a_0^2\frac{\omega_L^2}{\omega_p^2}$ \\
\hline
NL Lu  &$>2$   & $\frac{2\sqrt{a_0}}{k_p} $ & $\frac{4}{3}\frac{\sqrt{a_0}}{k_p}\frac{\omega_L^2}{\omega_p^2}$    &   $c\tau_L\frac{\omega_L^2}{\omega_p^2}$ &  $\sqrt{a_0}\frac{2\pi}{k_p}$ & $\sqrt{a_0}/2$ & $\frac{2}{3} a_0\frac{\omega_L^2}{\omega_p^2}$\\
\hline
NL GP &$>2\sqrt{\frac{n_c}{n_p}}$   & $\frac{\sqrt{a_0}}{k_p} $  & $ $    &   $a_0c\tau_L\frac{\omega_L^2}{\omega_p^2}$ &  $ $ & $\sqrt{a_0}$ & $ a_0^{\frac{3}{2}} \omega_p\tau_L\frac{\omega_L^2}{\omega_p^2}$ \\
\hline \hline
\end{tabular} 

\label{tab:scalinglaws}
\vspace{-0.02\textwidth}
\end{table} 
Both groups found optimum acceleration occurs when the laser pulse duration $\tau$ and spot size $w_b$ match the plasma wavelength, $w_b\sim c\tau\sim\sqrt{a_0}/k_p$, and the bubble resembles a spherical cavity with $r_b=w_0$, cf. Section \ref{ch:bubble_regime}.

\cite{bib:Pukhov2002} is based on a similarity theory with the parameter $S=\frac{n_{e}}{a_0n_c}$, valid for $S \ll 1$ and $a_0~>~4$. In this extreme case, the acceleration is limited by pump depletion, and the self-injected electrons posses a~quasi-monoenergetic spectrum. Its predictions for important quantities is given in Table \ref{tab:scalinglaws} labelled under NL GP. 
\cite{bib:Lu2007} uses a more phenomenological approach, with the evolution of the driver including etching effects (cf. Section \ref{ch:beamloading}), dephasing and beam loading, and is summarized in Table \ref{tab:scalinglaws} under "NL Lu". 
Its validity extends to lower laser intensities ($a_0 >2$) and is not restricted to 
self-injection\footnote{A comparison reveals that both scalings predit similar energy around several $\MeV{100}$'s to $\SI{1}{GeV}$ for  $n_{e,0}/n_{crit}=100-1000$ and $P=(10-100)TW$ and $c\tau_L \sim \lambda_p$. The accelerated charge is predicted to reach the \si{nC} regime - keeping in mind that \cite{bib:Pukhov2002} assumes a severely loaded wakefield, while \cite{bib:Lu2007} assumes an unperturbed wakefield}.
It shows good agreement with experimental results \cite{bib:McGuffey2012} and can also be applied for external injection schemes.
However, such scaling laws should be used with care, since they often overestimate experimental parameters.

In Figure \ref{fig:maximumGain} the maximum energy gain $W_{el}^{Lu}$ (Eq. (\ref{eq:max_energy_gain_3D})) is plotted with respect to $a_0$ and $n_{e,0}$. Again, the area below the white dotted line 
cannot be accessed without external guiding schemes. 
The~solid green line represents the matched beam condition ($w_b=2\sqrt{a_0}$) for the parameters given in the~caption, while the dashed blue line (top axis) gives the corresponding spot size $d_{FWHM}(a_0)$.
As long as the~self-guiding takes place, the maximum gain is achieved for low electron densities by avoiding dephasing effects. For example, for the same  laser parameters ($\SI{1.5}{J},\SI{28}{fs}$) the optimized case ($a_0=4.3$ and $n_{e,0}=\SI{5e18}{\per\cubic\cm}$) yields an energy gain of $W^{Lu}_{el}=\MeV{510}$, while $W_{el}^{Lu}=\MeV{380}$ at $a_0=6.3$ and $n_{e,0}=\SI{10e18}{\per\cubic\cm}$ .
\newline
\newline
\newline
\newline
\newline
\newline
\newline
\newline
\newline
\newline
\newline
\pagebreak
\section{Experimental scalings}

\begin{figure}[hbt!]
\includegraphics[width=\textwidth]{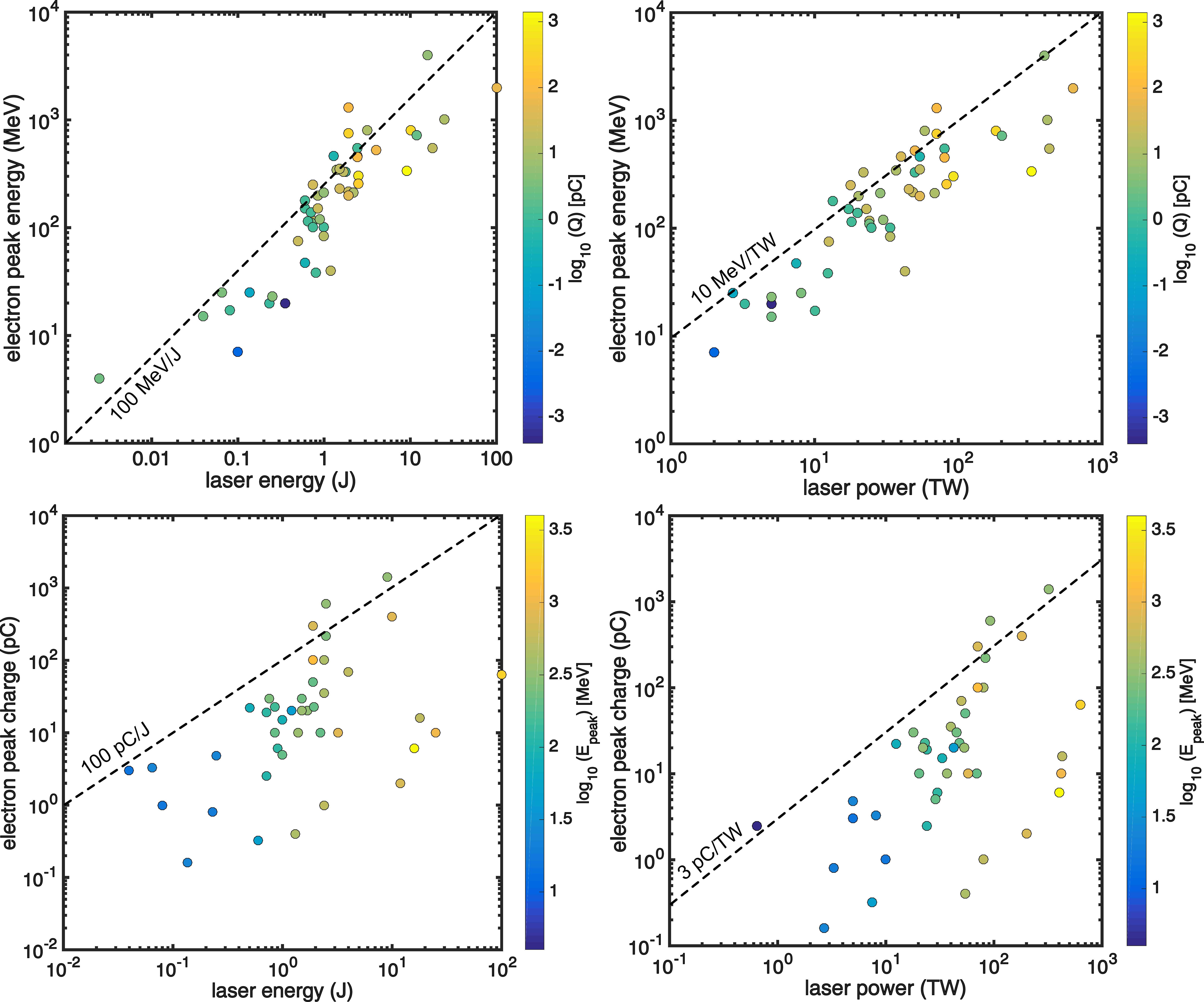}
\caption{\textbf{Experimental results for energy and charge}: Experimentally, the best results for electron peak energy and charge closely follow extremely simple scaling laws with respect to the laser power and energy. Note that these "laws" are no fit to the data, just lines to guide the eye. Data is based on 50+ publications on LWFA during the last 15-20 years \cite{bib:all_exp}}
\label{fig:expscaling}
\end{figure}

A remarkably simple pattern can be found for the experimental scaling of electron peak energy and charge with laser energy and power. In Figure \ref{fig:expscaling}, all combinations between of these quantities are plotted. Simple linear relations exist between the bunch energy or charge and the laser energy or power. This allows a straightforward order-of-magnitude perfomance prediction of LWFAs for a given laser. The dataset encompasses many different laser parameters, injection mechanisms, targets setups with or without guiding, but the linear behaviour seems to be maintained up to the PW level. The dataset focuses on quasi-monoenergetic spectra, and energies/charges are referring to the  spectral peak. 

In summary, the following scalings seem to hold for the upper end of the point clouds in Fig. \ref{fig:expscaling}:
\begin{align*}
O(E_{peak}) &\approx 100 \text{ MeV / J} \approx 10 \text{ MeV / TW}\\
O(Q_{peak}) &\approx 100 \text{ pC / J} \approx 3\text{ pC / TW}.
\end{align*}

\newpage

\section{Important equations}
\begin{table}[h]
\small
\renewcommand{\arraystretch}{1.6}
\centering
\caption{Important quantities for LWFA expressed in physical and engineering formula}
\begin{tabular}{|l|l|l|}\hline \hline
\textbf{ Quantity} & \textbf{Definition} & \textbf{Engineering Formula} \\

\hline
\hline
\multicolumn{3}{|c|}{ \bfseries \textit{Gaussian Laser Beam Parameters} \pmb{($a_0$)}}\\

\hline
Focal Spot & $2w_0\!=\!\frac{4\lambda_L}{\pi}\frac{f}{D}\!=\!\sqrt{\frac{2}{\ln 2}}d_{FWHM}$ & $ w_{\frac{1}{e^2}-\text{\O}}[\si{\micro\meter}]=f/\# \;\;\;\;\;\;\; \text{      @} \lambda_L=\SI{0.8}{\micro\meter}$ \\

\hline
Confocal Parameter & $2z_R=2\pi w_0^2/\lambda_L$ & $\Delta z[\si{\micro\meter}]=2(f/\#)^2\;\;\;\;\;\; \text{   @} \lambda_L=\SI{0.8}{\micro\meter}$\\
\hline
Peak Power & $P_0=2\sqrt{\frac{\ln 2}{\pi}}\frac{W_L}{t_{FWHM}}$   & $P_0[\si{\tera\watt}]=940\frac{W_L[\si{J}]}{t_{FWHM}[\si{fs}]}$ \\
& $P_0=\frac{\pi}{4\ln 2}d_{FWHM}^2I_0$ &$P_0[\si{\tera\watt}]=0.011d^2_{FWHM}[\si{\micro\meter}]I_0[10^{18}\frac{\si{W}}{\si{\square\centi\meter}}]$\\
\hline 
Peak Intensity & $I_0=\left(\frac{4\ln 2}{\pi}\right)^{\frac{3}{2}}\frac{W_L}{t_{FWHM}d^2_{FWHM}[\si{\micro\meter}]}$   &  $I_0[10^{18}\frac{\si{W}}{\si{\square\centi\meter}}]=83\times10^{3}\frac{W_L[\si{J}]}{t_{FWHM}[\si{fs}]d^2_{FWHM}[\si{\micro\meter}]}$ \\
&$I_0=\frac{2\pi^2\epsilon_0 m_e^2c^5}{e^2}\frac{a_0^2}{\lambda_L^2}$ & $I_0[10^{18}\frac{\si{W}}{\si{\square\centi\meter}}]=1.37\frac{a_0^2}{\lambda_L^2[\si{\micro\meter}]}$ \\
\hline

Vector Potential
& $a_0=\frac{e}{\pi m_e c^2}\sqrt{\frac{I_0}{2\epsilon_0c}}\lambda_L$ &$a_0=0.85\sqrt{I_0[\SI{e18}{\W\per\square\cm}]} \lambda_L[\si{\micro\meter}]$ \\
\hline

Peak Electric Field & $E_0=\frac{ea_0}{cm_e \omega_L}$ & $E_0[\SI{e12}{V\per\meter}]=3.2\frac{a_0}{\lambda_L[\si{\micro\meter}]}$\\
\hline


\hline
\multicolumn{3}{ |c|}{ \bfseries  \textit{ Plasma Parameters} \pmb{($n_e\propto k_p$)}}\\
\hline

Plasma Wavelength & $\omega_p=\sqrt{\frac{n_{e,0} e^2}{m_e \epsilon_0}}$ & $\lambda_p[\si{\micro\meter}]=\frac{33.4}{\sqrt{n_{e,0}[\SI{e18}{\per\cubic\cm}]}}$\\
\hline

Wavebreaking Field & $E_{p,0}=\frac{m_ec\omega_p}{e}$ & $E_{p,0}[\si{\giga\volt\per\m}]=96\sqrt{n_{e,0}[\SI{e18}{\per\cubic\cm}]}$\\
\hline

Plasma Gamma Factor & $\gamma_p=\sqrt{\frac{n_{cr}}{n_e}}$ & $\gamma_p=33.4\frac{1}{n_{e,0}[\SI{e18}{\per\cubic\cm}]\lambda_L[\si{\micro\meter}]}$\\
\hline

Critical Density & $n_{e,c}=\frac{\epsilon_0 m_e}{e^2} \omega_L^2$ & $n_{e,c}[$cm$^{-3}]=\frac{1.1 \times 10^{21}}{\lambda_L^2[\si{\micro\meter}]}$\\
\hline

\hline
\multicolumn{3}{|c|}{ \bfseries \textit{LWFA Parameters in the Bubble Regime \pmb{($r_b=2\sqrt{a_0}/k_p$)}}}\\
\hline

Dephasing Length & $L_d=\frac{2}{3\pi}\sqrt{a_0}\lambda_L\left(\frac{n_c}{n_{e,0}}\right)^{3/2} $ &  $L_{d}[\si{mm}]=7.9\sqrt{a_0}\left(\frac{\lambda_L^{-4/3}[\si{\micro\meter}]}{n_{e,0}[\SI{e18}{\per\cubic\cm}]}\right)^{{3}/{2}}$\\
\hline

Electric Field & $E_p\!\!=\!\!\frac{m_ec\omega_p}{e}\sqrt{a_0} $ & $E_{p}[\si{\giga\volt\per\m}]=96\sqrt{n_{e,0}[\SI{e18}{\per\cubic\cm}]}\sqrt{a_0}$\\
\hline

Electron Energy &  $ W_{el}=\frac{2a_0}{3}\left(\frac{n_c}{n_{e,0}}\right)m_ec^2 $ & $W_{el}[\si{\MeV}]\approx380\frac{a_0}{n_{e,0}[\SI{e18}{\per\cubic\cm}]\lambda_L^2[\si{\micro\meter}]}$\\
\hline

Optimum Charge & $Q_{opt}=\frac{\pi c^3}{e^2}\sqrt{\frac{m_e^3\epsilon_0^3}{n_{e,0}}}a_0^{\frac{3}{2}}  $ & $Q_{opt}[\si{\pico\coulomb}]=75\sqrt{\frac{a_0^3}{n_{e,0}[\SI{e18}{\per\cubic\cm}]}}$\\
\hline \hline 
\end{tabular}
\label{tab:quantities_LPA}
\end{table} 

\section{Summary}
We reviewed the basic physics of LWFA, starting from a description of the laser field of a Gaussian pulse, and the interaction of light with single atoms and electrons for extreme intensities, leading to the concept of the ponderomotive force as a repulsive net force of intense light on electrons. It leads to an expulsion of electrons from the laser axis, and the generation of moving plasma wave structure trailing the pulse as it propagates through a plasma medium. This so-called wakefield consists of plasma-electrons oscillating around their equilibrium position, a charge separation causing strong longitudinally accelerating and transversely focusing electric fields. We reviewed the wakefield generation in a 1-D analytical and 3D-phenomenological methods, and discussed the conditions and methods for trapping electrons in the accelerating regions of the wave. Finally, we discussed theoretical and experimental scaling laws for the process.


\end{document}